\documentclass[aps,pre,preprintnumbers,twocolumn,superscriptaddress,showkeys]{revtex4-2}

\usepackage{hyperref}
\usepackage{graphicx}
\usepackage{amsmath}
\usepackage{amssymb}
\usepackage{amsfonts}
\usepackage{upgreek}
\usepackage{color}
\usepackage{fourier}
\usepackage{bm}

\usepackage{xr}
\makeatletter

\newcommand*{\addFileDependency}[1]{
\typeout{(#1)}
\@addtofilelist{#1}
\IfFileExists{#1}{}{\typeout{No file #1.}}
}\makeatother

\newcommand*{\myexternaldocument}[1]{%
\externaldocument{#1}%
\addFileDependency{#1.tex}%
\addFileDependency{#1.aux}%
}

\myexternaldocument{arxiv-si-aux}

\begin{document}

\title{Thermodynamic uncertainty relation for systems with active Ornstein--Uhlenbeck particles}

\author{Hyeong-Tark Han}
\affiliation{{Department of Physics}, {POSTECH}, {{77 Cheongam-Ro}, {37673}, {Pohang}, {Republic of Korea}}}
\author{Jae Sung Lee}\email{jslee@kias.re.kr}
\affiliation{{School of Physics}, {Korea Institute for Advanced Study}, {{85 Hoegiro}, {02455}, {Seoul}, {Republic of Korea}}}
\author{Jae-Hyung Jeon}\email{jeonjh@postech.ac.kr}
\affiliation{{Department of Physics}, {POSTECH}, {{77 Cheongam-Ro}, {37673}, {Pohang}, {Republic of Korea}}}
\affiliation{{Asia Pacific Center for Theoretical Physics}, {{77 Cheongam-Ro}, {37673}, {Pohang}, {Republic of Korea}}}

\begin{abstract}
Thermodynamic uncertainty relations (TURs) delineate tradeoff relations between the thermodynamic cost and the magnitude of an observable’s fluctuation. While TURs have been established for various nonequilibrium systems, their applicability to systems influenced by active noise remains largely unexplored. Here, we present an explicit expression of TUR for systems with active Ornstein--Uhlenbeck particles (AOUPs). Our findings reveal that active noise introduces modifications to the terms associated with the thermodynamic cost in the TUR expression. The altered thermodynamic cost encompasses not only the conventional entropy production but also the energy consumption induced by the active noise. We examine the capability of this TUR as an accurate estimator of the extent of anomalous diffusion in systems with active noise driven by a constant force in free space. By introducing the concept of a contracted probability density function, we derive a steady-state TUR tailored to this system. Moreover, through the adoption of a new scaling parameter, we enhance and optimize the TUR bound further. Our results demonstrate that active noise tends to hinder accurate estimation of the anomalous diffusion extent. Our study offers a systematic approach for exploring the fluctuation nature of biological systems operating in active environments.
\end{abstract}

\maketitle

\section{Introduction}

Investigating the nature of fluctuations and their effects on dynamics is a central focus of research in stochastic thermodynamics. In equilibrium systems, fluctuations are connected to dissipation and the system's response to external perturbations, a relationship encapsulated by the fluctuation-dissipation theorem (FDT)~\cite{Pavliotis2014}. However, the FDT breaks down in nonequilibrium processes~\cite{HatanoSasa}, necessitating more generalized theoretical frameworks. In this context, fluctuation theorems~\cite{jarzynski1997nonequilibrium,crooks1999entropy} and various thermodynamic tradeoff inequalities~\cite{kwon2024unified}, such as thermodynamic uncertainty relations (TURs)~\cite{barato2015thermodynamic,gingrich2016dissipation,liu2020thermodynamic,koyuk2020thermodynamic,hasegawa2019uncertainty,lee2019TUR}, classical speed limits~\cite{Shiraish2018speedlimit,Dechant_2022Wasserstein,Kohei2021speed,lee2022speed}, entropic bound~\cite{EntropicBound}, and power-efficiency tradeoff relation~\cite{Benenti2011power,Shiraish2016power,pietzonka2018universal,lee2020power,proesmans2016power}, have emerged as crucial frameworks for systematically studying the impact of fluctuations in nonequilibrium processes.

Among the various tradeoff inequalities, TURs establish a connection between the variance of an observable $\Theta$ and the total entropy production (EP) $\Delta S_{\rm tot}$, expressed as follows~\cite{barato2015thermodynamic}:
\begin{equation}
\frac{\text{Var}[\Theta]}{\langle \Theta\rangle^2}\Delta S_\text{tot}\geq 2k_{\rm B},
\label{eq:tur}
\end{equation}
where $\text{Var}[\Theta]$ and $\langle \Theta\rangle$ represent the variance and mean value of the observable, respectively, and $k_{\rm B}$ denotes the Boltzmann constant. This relation quantifies a tradeoff between measurement precision affected by thermal fluctuations and the thermodynamic cost. Essentially, it reveals that achieving higher measurement precision requires increased thermodynamic cost, and vice versa. TURs can serve a dual purpose. Firstly, they provide a lower bound for the EP~\cite{Manikandan2020}. Researchers have used this bound to estimate EP and have further improved its accuracy by employing multidimensional observables~\cite{Dechant_multi_2019,Vu2020multi,DechantPRX2021} and entropic bound~\cite{SLee2023}. Secondly, TURs establish a lower limit for variance or uncertainty of an observable. Notably, recent work~\cite{hartich2021thermodynamic} demonstrates TUR's capability to {bound} the transition timescale between anomalous and normal diffusions. 

\begin{figure*}
\centering
\includegraphics[width=14cm]{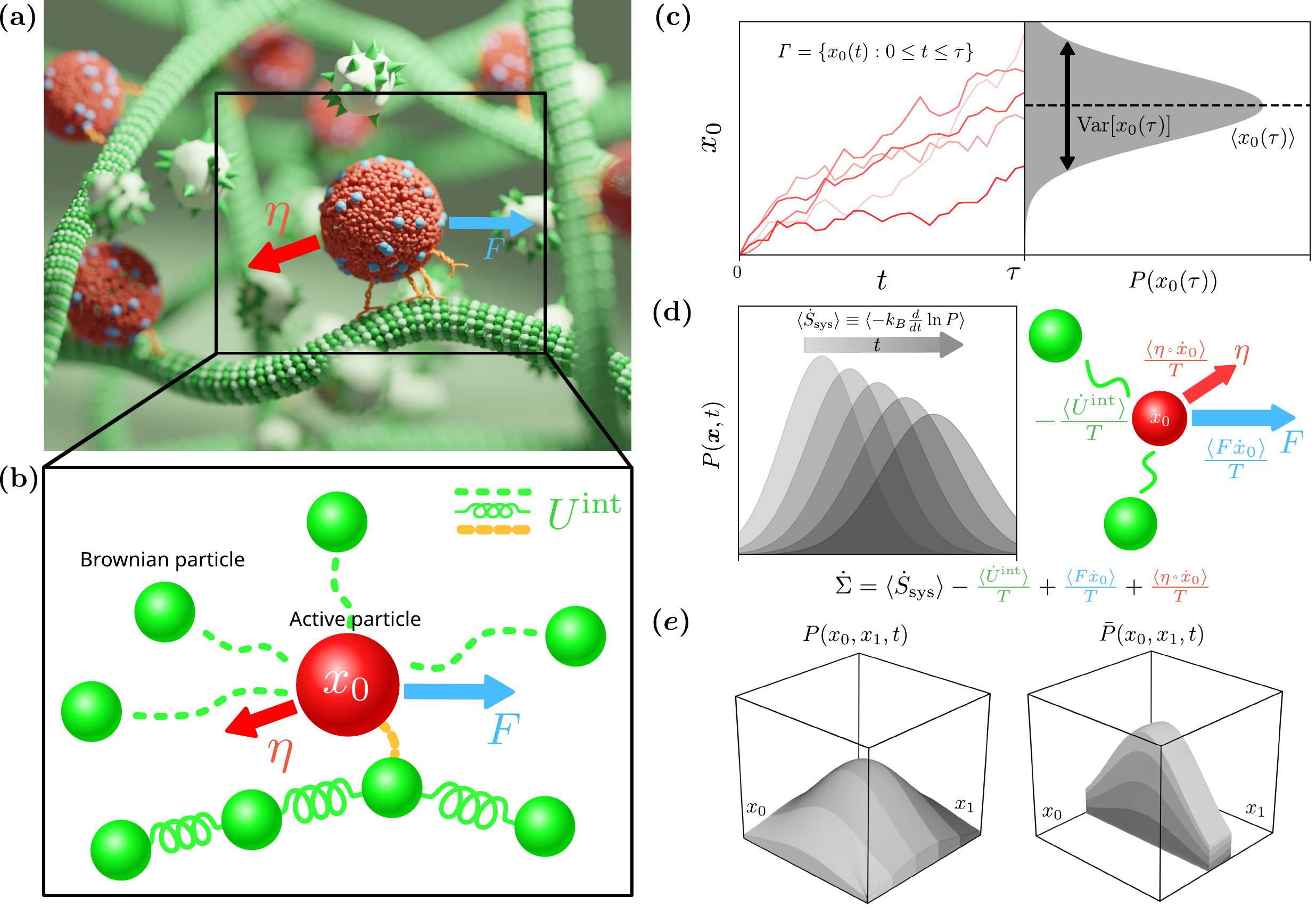}
\caption{ (a) An illustration of active transport by motor proteins on a microtubule. A group of kinesin and dynein motors exerts molecular forces on the vesicle, whose net effect results in a stochastic self-propulsive force ($\eta$).  Additionally, an external force ($F$), e.g., due to a flow or electric field, is applied to the vesicle. (b) The active transport system in (a) is simplified to a physical model depicted here: an active particle driven by a self-propulsive force ($\eta$) and an external force adheres to a polymer in a viscous environment containing Brownian particles. (c) The simulated trajectories ($\Gamma$) for these active particles in (b) and their probability density function [$P(x_0(\tau))$]. In the distribution, $\langle x_0(\tau)\rangle$ and $\mathrm{Var}[x_0(\tau)]$ are the mean position at $t=\tau$ and the variance, respectively. 
(d) The total dissipation rate ($\dot{\Sigma}$) and its components: The first term, $\langle \dot{S}_\mathrm{sys}\rangle$, explains the change of the Shannon entropy (Left panel). The other three terms are schematically represented in the Right panel. (e) The pdf $P$ for an AOUP in free space (Left) and its contracted pdf $\bar{P}$ (Right). In the latter, $x_0$ is bound to a finite domain in $0\leq x_0\leq L$.}
\label{fig1}
\end{figure*}

These examples highlight its applicability in biosystems {at the molecular level}, where the processes are inherently out of equilibrium~\cite{hopfield1974kinetic,murugan2012speed,skoge2013chemical,bialek2005physical,fisher1999force} and anomalous diffusion plays a pivotal role~\cite{barkai2012strange}. However, when applying TURs to biological systems, an additional distinctive feature of fluctuations must be considered: {small biological systems} are often subject to the influence of active noise~\cite{marchetti2013hydrodynamics,lauga2012dance,chen2015memoryless,harder2014activity,gal2013particle,wu2000particle,bechinger2016active}. This means that (small) biological entities frequently experience athermal noise from their environment~\cite{wu2000particle,bechinger2016active,samanta2016chain,nguyen2021active}, or exhibit self-propelled motion by consuming chemical energy~\cite{lauga2012dance,kafri2008steady,tailleur2008statistical,matthaus2009coli,ten2011brownian,romanczuk2012active,zheng2013non,howse2007self} (e.g., see Fig.~\ref{fig1}(a) \& (b)). As active noise fundamentally alters the characteristics of fluctuations, it becomes imperative to account for the active noise when formulating accurate thermodynamic relationships for such systems. Despite the significance of this aspect, the impact of active noise on TURs remains an area that has not been rigorously explored.

This study first aims to investigate the effect of active noise on TUR. For this purpose, we rigorously derive an explicit formulation of TUR via the information theoretic approach~\cite{hasegawa2019uncertainty} when a system comprises active Ornstein--Uhlenbeck particles (AOUPs)~\cite{wu2000particle,bechinger2016active,samanta2016chain,nguyen2021active}. We find that the presence of AOUPs modifies the term associated with the thermodynamic cost in TUR expression, caused by the energy consumption for sustaining self-propelled motion of AOUPs. 
We employ this modified TUR to estimate the lower bound of the mean square displacement (MSD) for two active systems (Fig.~\ref{fig1}(c)): a single AOUP and a Rouse chain consisting of an AOUP in an overdamped environment. Especially, we focus on the capability of the TUR to estimate the extent of anomalous diffusion when active noise is involved. Our analytic and numerical results demonstrate that active noise enhances fluctuation (or MSD) while simultaneously lowering the MSD bound due to the increment of energy consumption. In this sense, active noise hinders accurate estimation of the extent of anomalous diffusion using TUR. 

{Our TUR provides a foundational and practically useful framework for understanding the thermodynamic and dynamic properties of numerous biological systems.} Potential applications include analyzing the motion of loci or telomeres in chromosomes \cite{wang2008rapid,bronstein2009transient,bronshtein2015loss,stadler2017non,ku2022effects,yesbolatova2022formulation}, molecular transport in actin-myosin polymer networks \cite{harada1987sliding,amblard1996subdiffusion,wong2004anomalous,pollard2009actin}, and the dynamics of cross-linkers within endoplasmic reticulum networks \cite{vale1988formation,lin2014structure,speckner2018anomalous}. The modified TUR can be utilized to set thermodynamic bounds for the molecular motor-driven transport inside cells (Fig.~\ref{fig1}(a)) \cite{weihs2006bio,reverey2015superdiffusion,chen2015memoryless} or the motion of micro-swimmers in hydrogels \cite{lieleg2010characterization,caldara2012mucin,bej2022mucus}.

The rest of this paper is organized as follows. In the second section, we present our mathematical theory for the TUR of AOUPs embedded in an overdamped environment. Specifically, we introduce the model setup, terminology, and definitions used in this study. The derivation of the TUR is provided in the \textit{derivation of TUR} subsection.  In the following section, we elaborate on the application of the TUR to the lower bound of the MSD, focusing on a one-dimensional system driven by a constant force in free space. To resolve the problem of the absence of a steady state in free space, the \textit{steady state of contracted pdf and TUR} subsection introduces the concept of a contracted probability density function (pdf) and the corresponding steady-state TUR. The \textit{tighter bound} subsection introduces a new scaling factor for optimizing the TUR further. Concrete models, namely a single AOUP driven by a constant force and a Rouse chain consisting of an AOUP at its center, are examined in detail. 

\section{TUR of Active systems in Overdamped Environment}\label{sec:overdamped}

In this section, we begin by introducing the model systems, terminology, and definitions crucial for deriving the TUR of active systems in an overdamped environment. Subsequently, we present the TUR formula.

\subsection{Model Setup} \label{sec:model_overdamped}
Here, we present the formulation for a one-dimensional $N$-particle system in an overdamped environment. However, the extension to a $d$-dimensional system is straightforward. We denote the position of $i$th particle ($i=1,\cdots, N$) as $x_i$ and the active noise exerted on the $i$th particle as $\eta_i$. Using the notations $\bm{x}= (x_1, \cdots, x_N)^{\textsf T}$ and $\bm{\eta}= (\eta_1, \cdots, \eta_N)^{\textsf T}$, where $\textsf T$ denotes the matrix transpose, the equation of motion for the $i$th particle can be written as 
\begin{equation}
    \gamma \dot{x}_i = f_i ({\bm x},\bm{\eta},\omega t)+\eta_i+ \xi_i,
    \label{eq:eom_aoup}
\end{equation}
where $\gamma$ is the friction coefficient, $f_i$ represents the force exerted on the $i$th particle, $\omega$ is a parameter controlling the speed of a time-dependent protocol, and $\xi_i $ denotes a thermal Gaussian-white noise satisfying $\langle \xi_i (t)\rangle = 0$  and $\langle \xi_i(t) \xi_j(t^\prime)\rangle = 2 \gamma k_{\rm B} T \delta_{i,j} \delta(t-t^\prime)$. The force $f_i$ can be decomposed into two parts: an externally applied force, {which is generally space- and time-dependent, $f_i^{\rm ex}(\bm{x},\bm{\eta},t)$} and an interaction force among $N$ particles $-\partial_{x_i} U^{\rm int}$, i.e., $f_i = f_i^{\rm ex} -\partial_{x_i} U^{\rm int}$, where $U^{\rm int}=U^{\rm int} ({\bm x})$ denotes the interaction potential of the system. In this study, we adopt the AOUP noise for $\eta_i$, which phenomenologically simulates the self-propelled motion of an active particle or the noise from an athermal environment consisting of a bunch of active particles~\cite{bechinger2016active,wu2000particle,maggi2017memory}. Specifically, the dynamics of $\eta_i$ is governed by the following equation: 
\begin{equation} \label{eq:AOUPeq}
    \tau_{\rm A}\dot{\eta}_i = -\eta_i + \zeta_i,
\end{equation}
where $\tau_{\rm A}$ is called the persistence time and $\zeta_i$ is another Gaussian-white noise, {distinguished from the above thermal noise $\xi_i$}, satisfying
$\langle \zeta_i(t) \zeta_j(t')\rangle = 2\tau_A\gamma^2v_p^2\delta_{i,j}\delta(t-t')$. Here, $v_p$ can be interpreted as the propulsion speed of an active particle. Therefore, the AOUP noise $\eta_i$ is a Gaussian noise with zero mean and exponentially decaying correlation as $\langle \eta_i(t)\eta_j (t')\rangle = \gamma^2v_p^2 \delta_{i,j}\exp{(-|t-t'|/\tau_A)}$ in the steady state of $\eta_i$. 

For this active system, the Fokker--Planck equation governing the pdf $P(\bm{x},\bm{\eta},t;\omega)$, which includes the variables $x_i$ and $\eta_i$, is given by
\begin{equation}
   \partial_t P=\sum_i  (\mathcal{L}_{x_i} + \mathcal{L}_{\eta_i}) P,
   \label{eq:FP}
\end{equation}
where the operators $\mathcal{L}_{x_i}$ and $\mathcal{L}_{\eta_i}$ are defined as
\begin{equation}
    \mathcal{L}_{x_i} \equiv \partial_{x_i} \left(-\frac{g_i}{\gamma}+\frac{k_BT}{\gamma} \partial_{x_i} \right),~~~ \mathcal{L}_{\eta_i} \equiv
    \partial_{\eta_i} \left(\frac{\eta_i}{\tau_A}+\frac{\gamma^2v_p^2}{\tau_A} \partial_{\eta_i} \right) \label{eq:L_i}   
\end{equation}
with $g_i = g_i ({\bm{x}},\bm{\eta},t;\omega) \equiv f_i +\eta_i$. By defining the probability currents $J_{x_i}$ and $J_{\eta_i}$ as 
\begin{align} \label{eq:prob_curr_def}
    J_{x_i}  \equiv  \left(\frac{g_i}{\gamma} - \frac{k_BT}{\gamma} \partial_{x_i} \right) P ,~~~~
    J_{\eta_i}  \equiv -\left(\frac{\eta_i}{\tau_A}+\frac{\gamma^2v_p^2}{\tau_A} \partial_{\eta_i} \right) P
\end{align}
we can rewrite Eq.~\eqref{eq:FP} as $\partial_t P= - \sum_i \partial_{x_i} J_{x_i}- \sum_i \partial_{\eta_i} J_{\eta_i} $. 

In this overdamped environment, for a process starting at time $t=0$ and ending at time $t=\tau$, we are interested in  the following type of observable $\Theta_\tau (\omega)$:
\begin{equation} \label{eq:obs_over}
    \Theta_\tau (\omega)  = \int_0^\tau dt \; {\bm\Lambda}({\bm x}, {\bm \eta}, \omega t)^{\textsf T}  \circ \dot{{\bm x}} , 
\end{equation}
where ${\bm\Lambda}({\bm x}, {\bm \eta}, \omega t) = (\Lambda_1, \cdots, \Lambda_N)^{\textsf T}$ and $\circ$ denotes the Stratonovich product. From the identity $\langle {\bm H}^{\textsf T} \circ \dot{{\bm x}} \rangle = \int d{\bm x} d{\bm \eta} \; {\bm H}^{\textsf T}{\bm J}_{{\bm x}} $ for an arbitrary $N$-dimensional vector ${\bm H}$, the mean value of the observable is evaluated as 
\begin{equation} \label{eq:obs_avg}
\langle \Theta_\tau (\omega) \rangle  = \int_0^\tau dt\int d\bm{x}d\bm{\eta}\; {\bm\Lambda}({\bm x}, {\bm \eta}, \omega t)^{\textsf T} {\bm J_{\bm x}} , 
\end{equation}
where ${\bm J}_{\bm x} = (J_{x_1}, \cdots, J_{x_N})^{\textsf T}$. 

\subsection{Derivation of TUR} \label{sec:derivation_overdamped}

The derivation of the TUR is based on the method using the Cram\'er--Rao inequality along with perturbed dynamics~\cite{hasegawa2019uncertainty}. We consider the following Fokker--Planck equation, achieved by multiplying the perturbation factor $1+\theta$ to the operator $\mathcal{L}_{x_i} + \mathcal{L}_{\eta_i}$ in Eq.~\eqref{eq:FP} as
\begin{equation} \label{eq:perturbedFP}
    \partial_t P_\theta=  (1+\theta)\sum_i (\mathcal{L}_{x_i} + \mathcal{L}_{\eta_i}) P_\theta,
\end{equation}
where $P_\theta = P_\theta (\bm{x},\bm{\eta},t;\omega)$ is the pdf of the perturbed dynamics. This $\theta$-perturbed dynamics returns back to the original dynamics at $\theta=0$. $P_\theta$ is related to the pdf of the original dynamics $P$ in Eq.~\eqref{eq:FP} through
\begin{equation}
P_\theta (\bm{x},\bm{\eta},t;\omega)= P(\bm{x},\bm{\eta},t_\theta;\omega_\theta),
\end{equation}
where $t_\theta\equiv (1+\theta)t$ and $\omega_\theta\equiv \omega/(1+\theta)$. 
This perturbation scheme enables us to derive the TUR by using the Cram\'er--Rao inequality~\cite{hasegawa2019uncertainty}
\begin{equation} \label{eq:CramerRao}
    \frac{[\partial_\theta \langle \Theta_\tau(\omega) \rangle_\theta]^2}{{\rm Var}_\theta [\Theta_\tau (\omega)]} \leq \mathcal{I}(\theta), 
\end{equation}
where ${\rm Var}_\theta [\cdots]$ represents the variance of an observable and $\mathcal{I}(\theta) \equiv \langle -\partial_\theta^2 \ln \mathcal{P}_\theta (\Gamma) \rangle_\theta$ denotes the Fisher information with respect to the path probability $\mathcal{P}_\theta (\Gamma)$ of a trajectory $\Gamma$ in the $\theta$-perturbed dynamics. 

Evaluation of the Cram\'er--Rao inequality at $\theta = 0$ leads to the TUR. To do so, we derive the analytic expressions for $\partial_\theta \langle \Theta_\tau(\omega) \rangle_\theta$ and the Fisher information at $\theta = 0$ from our active Langevin model (see the Supplementary Section~\ref{secA:detailed_derivation} for details). 
By combining these results with the Cram\'er--Rao inequality, we arrive at the general form of the TUR for the AOUP system:
\begin{equation} \label{eq:activeTUR}
    \frac{{\rm Var}[\Theta_\tau (\omega)]}{[\hat{h}\langle \Theta_\tau (\omega) \rangle]^2} \Sigma_\tau \geq 2 k_{\rm B},
\end{equation}
which represents the first main result of our work. Here, the operator $\hat h$ is defined as $\hat{h} \equiv \tau\partial_\tau - \omega \partial_\omega$, and  $\Sigma_\tau$ is the thermodynamic cost
\begin{equation}\label{eq:therdmodynamiccost}
    \Sigma_\tau=\int_0^\tau dt \int dx d\eta \sum_i \left(\frac{\gamma J_{x_i}^2}{TP}+\frac{k_B\tau_A J_{\eta_i}^2}{\gamma^2 v_p^2 P}\right)
\end{equation}
derived from the Fisher information $\mathcal{I}(0)$ and the relation $\mathcal{I}(0) = \frac{1}{2k_{\rm B}} \Sigma_\tau $~\eqref{eq:Fisher1}.
{Here, it is important to note that there is still an ambiguity in the thermodynamically consistent entropy production that generally holds for all active systems~\cite{pietzonka2017entropy,padmanabha2023fluctuations, fritz2023thermodynamically}. However, in our system the active noise is a mechanical force attributed to the activity of a certain physical entity, such as a chemical driving force of the Janus particle. Accordingly, we interpret each term in this framework as follows.}
The first term on the right-hand side of Eq.~\eqref{eq:therdmodynamiccost} is the conventional EP known for the ordinary Langevin system, while the second term explains the additional energy consumption from the active dynamics. Corroborating this interpretation, we further identify that 
\begin{align}\label{eq:EPexpression}
    \dot{\Sigma}_\tau&=\langle \dot{S}_{\rm sys} \rangle -\sum_i  \frac{\langle \dot{Q}_i^f\rangle}{T} + \sum_i \frac{\langle \eta_i \circ \dot{x}_i \rangle}{T}
\end{align}
(the derivation is shown in Supplementary Section \ref{secA:sigma_insight}). 
As summarized in Fig.~\ref{fig1}(d), the thermodynamic cost comprises two components as $\Sigma_\tau= \Sigma_\tau^{\rm con} + \Sigma_\tau^{\rm act}$, i.e., the conventional EP $\Sigma_\tau^{\rm con} \equiv \langle \dot{S}_{\rm sys} \rangle -\sum_i \langle \dot{Q}_i^f\rangle/T$ and the energy consumption by the active motion $\Sigma_\tau^{\rm act} \equiv \sum_i \langle \eta_i \circ \dot{x}_i \rangle/T$. The new TUR~\eqref{eq:activeTUR} returns back to the previous one~\cite{koyuk2020thermodynamic} when the active noise disappears.

\section{Application to the bound on mean square displacement} \label{sec:applicationToMSD}

Inspired by the biological active transport exemplified in Fig.~\ref{fig1}(a), we now focus on an AOUP system dragged by a constant force and investigate the bound on its MSD using TUR. We consider a one-dimensional system consisting of $N=2M+1$ particles, where the particle index $i$ ranges from $-M$ to $M$. The constant force $F$ and AOUP noise $\eta$ are applied only to the center particle with index $i=0$. The interaction potential between the particles, denoted by $U^{\rm int} (\bm x)$, is assumed to bind the system as a whole, preventing particles from being far away from each other, and to be a function of distance between any two particles. Then, from Eqs.~\eqref{eq:eom_aoup} and \eqref{eq:AOUPeq}, the equation of motion governing this system can be written as
\begin{align} \label{eq:MSDeom}
    &\gamma \dot{x}_i = -\partial_{x_i} U^{\rm int}({\bm x}) + \delta_{i,0} (F+\eta) + \xi_i, \nonumber \\
    & \tau_{\rm A} \dot{\eta} = -\eta +\zeta. 
\end{align}
Note that the index for the active noise is not necessary in this example since it is exerted only on a single particle. 
The corresponding Fokker-Planck equation of the pdf $P({\bm x},\eta,t) = P(x_0,x_{\pm 1}, \cdots, x_{\pm M},\eta,t) $ is 
\begin{equation} \label{eq:MSD_FP}
    \partial_t P = -\sum_{i=-M}^M \partial_{x_i} \hat{J}_{x_i} P - \partial_{\eta} \hat{J}_{\eta} P,
\end{equation}
where $\hat{J}_{x_i} \equiv [-\partial_{x_i} U^{\rm int}(\bm x) +\delta_{i,0} (F+\eta) - k_{\rm B}T \partial_{x_i}]/\gamma$ and $\hat{J}_\eta \equiv (-\eta-\gamma^2 v_p^2 \partial_\eta)/\tau_{\rm A}$.

For investigating MSD, we take the observable as the displacement of the center active particle during time lag $\tau$, that is, 
\begin{equation} \label{eq:MSD_obs}
\Theta_\tau = \int_0^\tau dt \; \dot{x}_0 = x_0(\tau) - x_0 (0) \equiv \Delta x_0(\tau)  
\end{equation}
by choosing $ \Lambda_i = \delta_{0,i}$. After a certain transition period, the system reaches the state where it moves in a constant velocity $\langle \dot{x}_i \rangle^{\rm ss} = v^{\rm ss}$. We can evaluate $v^{\rm ss}$ by taking average of the upper equation in Eq.~\eqref{eq:MSDeom} and summing it over all $i$, which results in $(2M+1)\gamma v^{\rm ss} = -\langle \sum_i \partial_{x_i} U^{\rm int} ({\bm x}) \rangle + F$. Applying the action-reaction law $\sum_i \partial_{x_i} U^{\rm int} ({\bm x}) = 0$ leads to 
\begin{equation}
    v^{\rm ss} = F/ (N\gamma).  
\end{equation}
It is noted that $v^{\rm ss}$ remains a constant regardless of both the form of $U^{\rm int}({\bm x})$ and the active noise. We will study the lower bound of MSD  using TUR in the constant-velocity state. Importantly, the constant-velocity state cannot be simply referred to as the steady state of this process. In fact, there is no steady state in this process, as the dynamics are essentially identical to the diffusive motion in free space. {\color{black}Thus, the application of the steady-state TUR to this process necessitates special treatment.} To address this issue, we introduce the steady state of the \emph{contracted pdf} in the subsequent section.

\subsection{Steady state of contracted pdf and TUR} \label{sec:contractedPDF}

We consider the Langevin dynamics described by Eq.~\eqref{eq:MSDeom}, but with the origin being varied depending on $x_0(t)$: the origin is shifted to $n L$ when $nL \leq x_0(t) < (n+1)L$ in terms of the unchanged coordinate, where $n$ is an integer and $L$ denotes an arbitrary positive constant in length unit. In the shifted coordinate, $x_0$ is confined within the range $0 \leq x_0 \leq L$, while the others $x_i$ $(i\neq 0)$ have no such restriction {and diffuse freely in an infinite space, distinguishing the description from the one on a ring}. Then, the pdf at the position $(x_0,x_{\pm 1}, \cdots, x_{\pm M})$ in the shifted coordinate is given by the sum over pdfs at all $(x_0 +nL,x_{\pm 1} +nL, \cdots, x_{\pm M}+ nL)$ positions in the unchanged coordinate ($n\in \mathbb{Z}$). That is, the pdf of the shifted coordinate $\bar{P}$ can be written in the following contracted form: 
\begin{align} \label{eq:contracted_pdf}
    &\bar{P}(x_0,x_{\pm 1}, \cdots, x_{\pm M},\eta,t) \nonumber \\
    &\equiv \sum_{n = -\infty}^\infty P(x_0 +nL,x_{\pm 1} +nL, \cdots, x_{\pm M}+ nL,\eta,t)
\end{align}
where $0\leq x_0\leq L$ in the contracted space.  Fig.~\ref{fig1}(e) visualizes the profiles of the original pdf and its contracted form $\bar{P}$. We can easily verify that (i) $\bar{P}$ is normalized over the contracted space, i.e., $\int d{\bm x} d\eta \bar{P} =1$ and (ii) $\bar{P}$ is the solution of the original Fokker--Planck equation~\eqref{eq:MSD_FP} with the boundary condition $\bar{P}(0,x_{\pm 1}, \cdots, x_{\pm M},\eta,t) = \bar{P}(L,x_{\pm 1}+L, \cdots, x_{\pm M}+L,\eta,t)$. As the variable $x_0$ is limited within a finite domain and all other particles are bound via the interaction potential, there exists the steady-state pdf $\bar{P}^{\rm ss}$.

Under this setup, we consider the Fokker--Planck equation for the $\theta$-perturbed dynamics similar to Eq.~\eqref{eq:perturbedFP} as
\begin{equation} \label{eq:perturbed_MSD_FP}
    \partial_t \bar{P}_\theta  = -(1+\theta)\left[\sum_{i=-M}^M \partial_{x_i} \hat{J}_{x_i} \bar{P}_\theta + \partial_{\eta} \hat{J}_{\eta} \bar{P}_\theta \right].
\end{equation}
It is straightforward to see that $\bar{P}_\theta (x_0,\cdots,x_{\pm M},\eta,t) = \bar{P} (x_0,\cdots,x_{\pm M},\eta,t_\theta)$, thus, $\bar{P}_\theta^{\rm ss} = \bar{P}^{\rm ss}$.
As the observable is the displacement of the particle with index $0$ in the unchanged coordinate, the variance of the observable remains unchanged. Furthermore, we can evaluate the steady-state mean observable in the perturbed dynamics as 
\begin{align} \label{eq:pert_obs_ss}
    \langle \Theta_\tau \rangle_\theta^{\rm ss} &= \int_0^\tau dt \langle \dot{x}_0 \rangle_\theta^{\rm ss} = \int_0^\tau dt \int_{\rm cont} d{\bm x}d\eta (1+\theta)\hat{J}_{x_0} \bar{P}^{\rm ss}  \nonumber \\ &= (1+\theta) \int_0^\tau dt \int_{\rm all} d{\bm x} d\eta \hat{J}_{x_0} P = (1+\theta) v^{\rm ss} \tau,
\end{align}
where $\int_{\rm cont} d{\bm x} d\eta \equiv \int_0^{L} d x_0 \int_{-\infty}^{\infty} \prod_{i\neq 0} dx_i d\eta$ denotes integration over a contracted space and $\int_{\rm all} d{\bm x} d\eta \equiv \int_{-\infty}^{\infty} \prod_{{\rm all~} i} d x_i d\eta$ represents integration over all space. We can verify the third equality in Eq.~\eqref{eq:pert_obs_ss} by substituting $\bar{P}$ with Eq.~\eqref{eq:contracted_pdf}. The fourth equality in Eq.~\eqref{eq:pert_obs_ss} comes from the relation $\langle \dot{x}_i \rangle 
 = \int d{\bm x} d\eta \hat{J}_{x_i} P$. 
The Fisher information of the contracted pdf is given as
\begin{align} \label{eq:Fisher_MSD_contracted}
\mathcal{I}(0)= \frac{1}{2k_{\rm B}} \left(\frac{F v^{\rm ss} \tau}{ T} + \frac{ \langle \eta  \circ \dot{x}_0 \rangle^{\rm ss} \tau}{T}\right) = \frac{\dot{\Sigma}^{\rm ss}\tau}{2 k_{\rm B}},
\end{align}
where $\dot{\Sigma}^{\rm ss} \equiv (Fv^{\rm ss} + \langle \eta \circ \dot{x}_0\rangle^{\rm ss})/T$ represents the steady-state thermodynamic cost in the contracted space (See Supplementary Section \ref{secA:FisherInformation_contractedpdf} for details).  Plugging Eqs.~\eqref{eq:pert_obs_ss} and ~\eqref{eq:Fisher_MSD_contracted} into the Cram\'er--Rao inequality~\eqref{eq:CramerRao} results in the TUR as
\begin{equation} \label{eq:contracted_ssTUR}
    Q \equiv \frac{{\rm Var}[\Delta x_0(\tau)]}{F^2N^{-2}\gamma^{-2} \tau} \dot{\Sigma}^{\rm ss} \geq 2k_{\rm B} .
\end{equation}
In fact, this steady-state TUR~\eqref{eq:contracted_ssTUR} can be obtained in an ad hoc manner using the original $P$ and the TUR \eqref{eq:activeTUR} by simply omitting the non-steady contribution $\langle \dot{S}_{\rm sys}\rangle$ in Eq.~\eqref{eq:EPexpression}.
However, this is not correct because the steady state does not exist in the original free diffusion process.
{Also note that the current theoretical framework can also be extended to the active system under periodic forces that satisfy $f_i (x + L) = f_i (x)$. }

\subsection{Tighter bound} \label{sec:tighterbound}

The TUR~\eqref{eq:contracted_ssTUR} for the displacement can be further optimized. For this purpose, we extend the Langevin equation~\eqref{eq:MSDeom} with the introduction of a new scaling parameter $h$ and the optimization factor $\alpha$ as
\begin{align} \label{eq:MSDeom_h}
    &\gamma \dot{x}_i = -\partial_{x_i} U^{\rm int}({\bm x}) + \delta_{i,0} [F+(1-\alpha)\eta] + \delta_{i,0} h \alpha \eta  + \xi_i, \nonumber \\
    & \tau_{\rm A} \dot{\eta} = -\eta +\zeta. 
\end{align}
Note that Eq.~\eqref{eq:MSDeom_h} returns back to Eq.~\eqref{eq:MSDeom} when $h=1$. From the derivation presented in Supplementary Section~\ref{sec:tighter_bound}, we obtain the steady-state TUR at $h=1$ as follows:
\begin{align} \label{eq:tighter_TUR}
    \frac{{\rm Var}[\Delta x_0(\tau)]}{F^2N^{-2}\gamma^{-2} \tau} \dot{\Sigma}^{\rm ss}(\alpha) \geq 2 k_{\rm B},   
\end{align}
where $\dot{\Sigma}^{\rm ss}(\alpha)$ is defined as
\begin{align} \label{eq:Sigma_alpha}
    \dot{\Sigma}^{\rm ss} (\alpha) \equiv 
    \frac{Fv^{\rm ss}}{T} + \frac{1-2\alpha}{T} \langle \eta \circ \dot{x}_0\rangle^{\rm ss} + \frac{ \alpha^2}{T\gamma} \langle \eta^2\rangle^{\rm ss}.
\end{align}
$\dot{\Sigma}^{\rm ss}(\alpha)$ is minimized at $\alpha^* \equiv \langle \eta\circ \dot{x}_0 \rangle^{\rm ss} /(\gamma v_p^2)$ which is the solution of  $\partial_\alpha \dot{\Sigma}^{\rm ss}(\alpha) = 0$. When $\alpha\to0$, Eq.~\eqref{eq:Sigma_alpha} converges to the $\dot{\Sigma}^{\rm ss}$ defined in the TUR \eqref{eq:contracted_ssTUR}. Putting $\alpha^*$ into $\dot{\Sigma}^{\rm ss} (\alpha)$ results in the {\color{black}optimized} TUR as follows:
\begin{align} \label{eq:tighterQ}
    Q^* \equiv \frac{{\rm Var}[\Delta x_0 (\tau)]}{{v^{\rm ss}}^2 \tau} \dot{\Sigma}^{\rm ss}(\alpha^*) \geq 2 k_{\rm B}, 
\end{align}
where $\dot{\Sigma}^{\rm ss} (\alpha^*) = (Fv^{\rm ss} + \langle \eta \circ \dot{x}_0\rangle^{\rm ss})/T-[\langle \eta\circ \dot{x}_0 \rangle^{\rm ss}]^2/(\gamma v_p^2T)$. This {\color{black}tighter TUR for translational symmetrical AOUPs} is the second main result of our study.
{Note that the optimization of the TUR bound can be understood as a mathematical technique for maximizing the bound in the absence of in priori information of $\alpha$ values. }

\begin{figure}
    \centering
    \includegraphics[width=8cm]{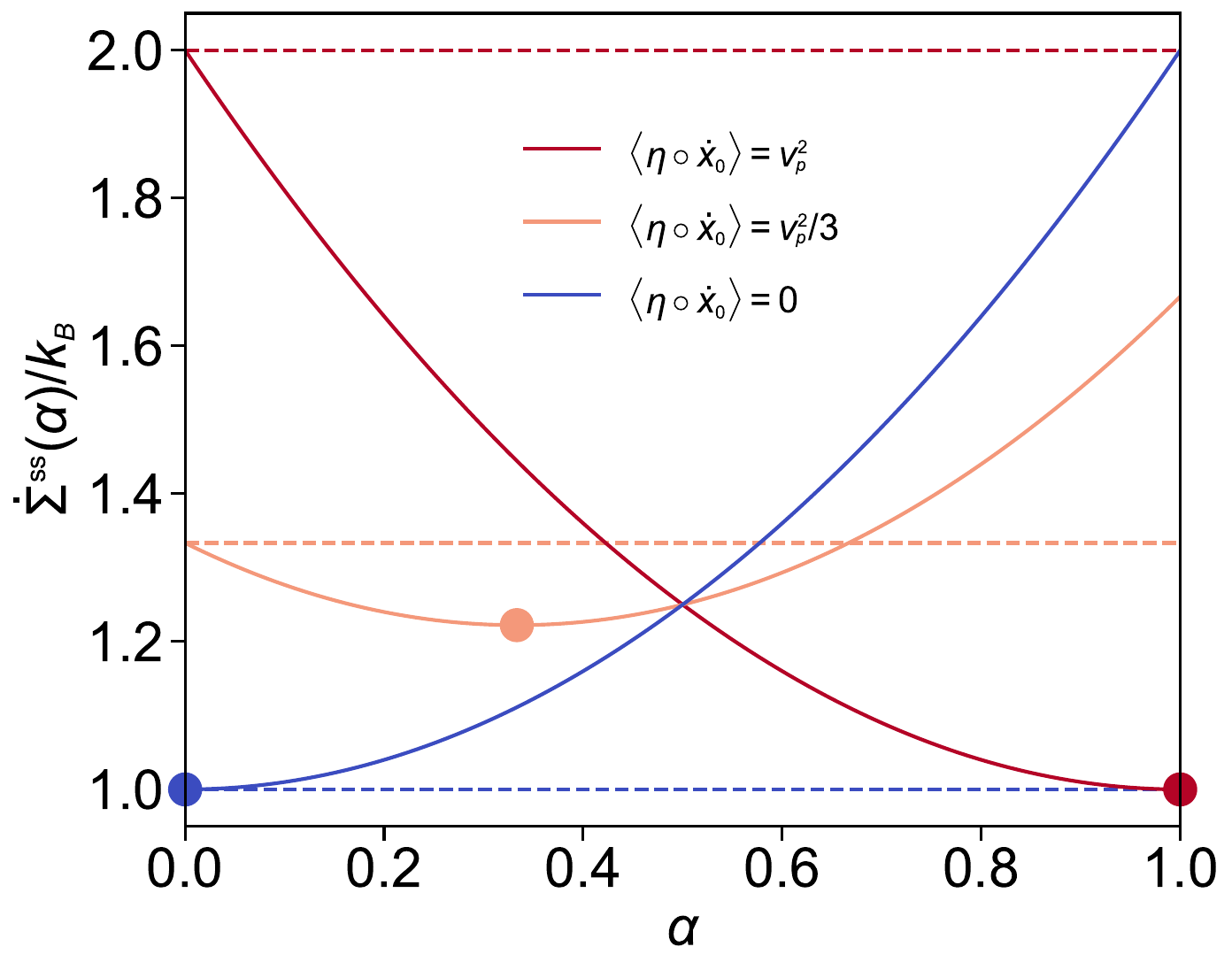}
\caption{Profiles of $\dot\Sigma^{\rm ss}(\alpha)/k_\mathrm{B}$ [Eq.~\eqref{eq:Sigma_alpha}] as a function of $\alpha$ for three distinct cases. In this numerical plot, we choose $Fv^\mathrm{ss}=\gamma v_p^2=1~k_\mathrm{B}T$. While the solid line represents $\dot\Sigma^{\rm ss}(\alpha)/k_\mathrm{B}$, the dashed line indicates the zeroth value before optimization, $\frac{F v^{\rm ss} }{ k_\mathrm{B}T} + \frac{ \langle \eta  \circ \dot{x}_0 \rangle^{\rm ss} }{k_\mathrm{B}T}(=\dot\Sigma^{\rm ss}(\alpha=0)/k_\mathrm{B})$, defined in Eq.~\eqref{eq:Fisher_MSD_contracted}. The circles mark the position of the $\alpha^*$.}
\label{fig2}
\end{figure}
In Fig.~\ref{fig2}, we numerically plot $\dot\Sigma^{\rm ss}(\alpha)/k_\mathrm{B}$ (the solid line) as a function $\alpha$ for three cases of $\langle \eta\circ \dot x_0\rangle^{\rm ss}$. 
In the plot, the circles indicate the position of the $\alpha^*$, and the dashed lines represent the zeroth value of $\dot\Sigma^{\rm ss}/k_\mathrm{B}$ at $\alpha\to0$ defined in  Eq.~\eqref{eq:Fisher_MSD_contracted}.
For active soft matter systems~\cite{caspi2000enhanced,sprakel2007rouse,han2023nonequilibrium}, the dissipation rate from $\eta$ lies in the range of $0<\langle \eta\circ \dot x_0\rangle^{\rm ss}<v_p^2$ due to the viscoelastic feedback (as shown in the example of the active Rouse chain (Sec.~\ref{sec:Rouse}) and Eq.~\eqref{eq:eta_dotx_expression}). In this case (e.g., the yellow solid line), $\dot\Sigma^{\rm ss}(\alpha)/k_\mathrm{B}$ is of the convex shape and minimized at the value of $0<\alpha^*<1$, which leads to a tighter TUR. However, for the passive system ($\langle  \eta \circ \dot x_0 \rangle = 0$, blue), the minimum point of $\dot\Sigma^{\rm ss}(\alpha)/k_\mathrm{B}$ exists at $\alpha=0$, so the TUR is already tight and further optimization procedure is unnecessary.   
Notably, at the maximal active dissipation rate ($\langle  \eta \circ \dot x_0 \rangle/v_p^2 =1$, red),  $\dot\Sigma^{\rm ss}(\alpha)/k_\mathrm{B}$ is monotonically decreasing with $\alpha$, where $\alpha^*=1$ and $\dot\Sigma^{\rm ss}(\alpha^*)$ becomes identical to the $\dot\Sigma^{\rm ss}$ for the passive system. The optimization is more effective for active systems with larger $\alpha^*$. The physical meaning of the $\alpha$ value pertains to the time reversibility of the active noise, as discussed in detail in Sec.~\ref{sec:conclusion}.

\subsection{Case 1: Single AOUP driven by a constant force} \label{sec:singleAOUP}

\begin{figure}
\centering
\includegraphics[width=8cm]{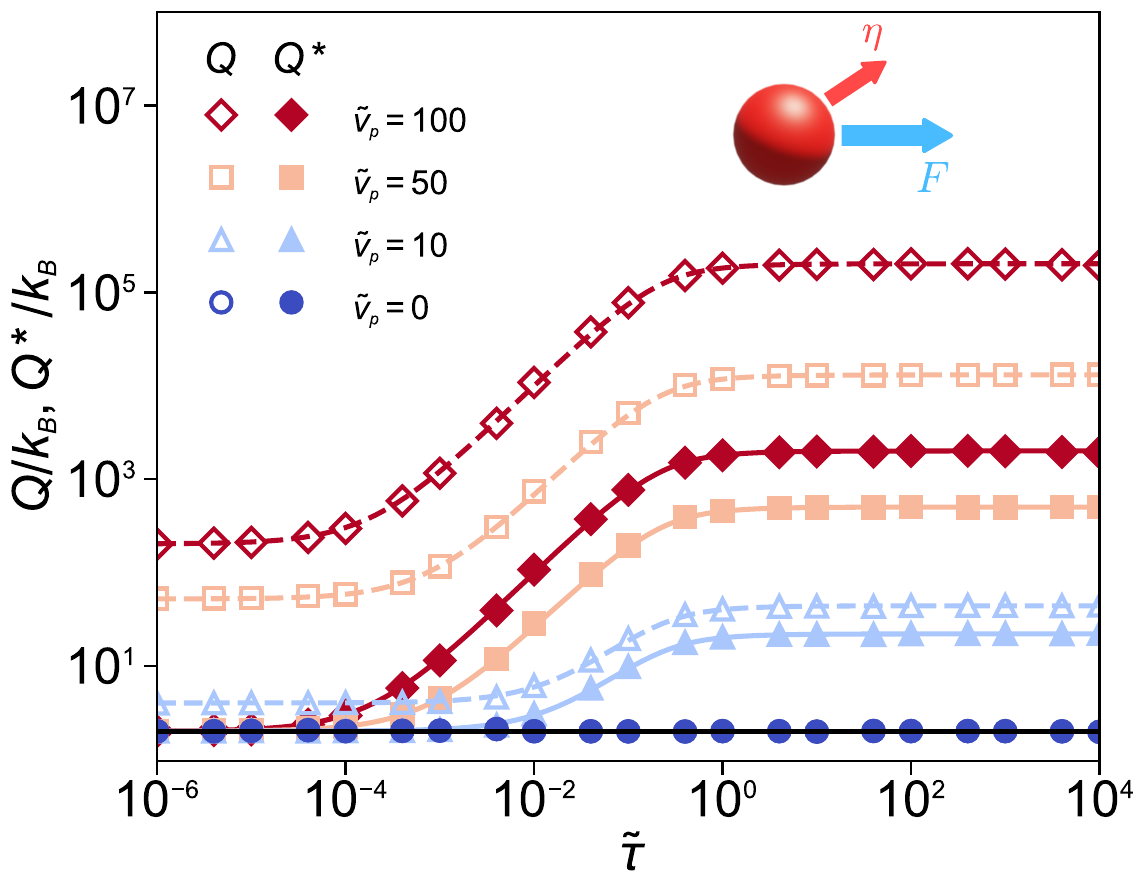}
\caption{ Plots of $Q/k_{\rm B}$ [Eq.~\eqref{eq:contracted_ssTUR}] and $Q^*/k_{\rm B}$ [Eq.~\eqref{eq:tighterQ}] as a function of dimensionless time $\tilde \tau$ for a single AOUP driven by a constant force. 
Empty and filled markers represent the $Q/k_B$ and $Q^*/k_B$, respectively, for various P\'eclet numbers $\tilde v_p = 0, 10, 50$, and $100$ with a given $\tilde \tau_A=0.1$. 
The colored dashed and solid curves denote the analytic expressions of $Q/k_B$ and $Q^*/k_B$, respectively. The black horizontal solid line illustrates the lower limit of the TUR, i.e., $2$. 
}
\label{fig3}
\end{figure}

As the simplest example, we consider a single AOUP by a constant force $F$ in one-dimensional space driven  (the inset of Fig.~\ref{fig3}). Its equation of motion is described by Eq.~\eqref{eq:MSDeom} with $M=0$ and $U^{\rm int} =0$. By setting $t_0$ and $\ell_0 \equiv \sqrt{k_{\rm B}T t_0/\gamma} $ as time and length unit, respectively, we can rewrite the equation of motion in a dimensionless form with $\tilde{x} \equiv x/\ell_0$ and $\tilde t \equiv t/t_0$, redefining $\tilde{F} = Ft_0/(\gamma \ell_0)$, $\tilde v_p = v_p t_0/\ell_0$, and $\tilde{\tau}_{\rm A} = \tau_{\rm A}/t_0$.
This dimensionless formulation amounts to setting $\gamma =1$ and $k_{\rm B} T=1$ in the original equation of motion. 
The parameter values used for the numerical simulations presented in this and the next subsections are $\tilde \tau_{\rm A} = 0.1$ and $\tilde F = 10$. The $\tilde v_p$ can be considered as the P\'eclet number defined by $\ell_0 v_p/ D$ which is the ratio between the active noise strength $\ell_0 v_p$ and the diffusivity $D\equiv k_{\rm B}T/\gamma$. 

The mean velocity of this system in the steady state is $\langle \frac{d \tilde x}{d\tilde t }\rangle^{\rm ss} = v^{\rm ss} t_0/\ell_0 = \tilde F $. Moreover, in the steady state, we can show that the MSD or the variance of the displacement becomes
\begin{equation} \label{eq:MSD_singleAOUP}
\text{Var}[\Delta \tilde x(\tilde \tau)] = 2 \tilde \tau +2 \tilde v_p^2 \tilde\tau_A \left[ \tilde \tau -\tilde \tau_{\rm A}( 1- e^{-\tilde \tau/\tilde \tau_A}) \right].
\end{equation}
The detailed derivation of Eq.~\eqref{eq:MSD_singleAOUP} is presented in Supplementary Section \ref{sec:MSD_singleAOUP}. Finally, $\langle \tilde \eta \circ \dot{\tilde x} \rangle^{\rm ss}$ can be calculated as
\begin{align} \label{eq:eta_dotx_singleAOUP}
    \langle \tilde \eta \circ \dot {\tilde x} \rangle^{\rm ss} = F \langle \tilde \eta\rangle + \langle \tilde \eta^2\rangle + \langle \tilde \eta \circ \tilde \xi\rangle = \tilde v_p^2,
\end{align}
where $\langle \tilde \eta \rangle = \langle \tilde \eta \circ \tilde \xi\rangle =0$ is used. Inserting Eqs.~\eqref{eq:MSD_singleAOUP} \&~\eqref{eq:eta_dotx_singleAOUP} into  Eq.~\eqref{eq:contracted_ssTUR} yields the $Q$ factor at time $\tilde\tau$ as follows:
\begin{equation}
\frac{Q(\tilde \tau)}{k_{\rm B}}=2\bigg[1 + \frac{\tilde v_p^2 \tilde \tau_A^2}{ \tilde \tau}\Big(e^{-\tilde \tau/\tilde \tau_A}-1+\frac{\tilde \tau}{\tilde \tau_A}\Big)\bigg]
\bigg[1+\Big(\frac{\tilde v_p}{\tilde F}\Big)^2\bigg]. 
\label{eq:q_aoup_thm}
\end{equation}
From Eq.~\eqref{eq:tighterQ}, we can also obtain the tighter bound $Q^*$ as
\begin{equation}
\frac{Q^* (\tilde \tau)}{k_{\rm B}}= 2\left[ 1 + \frac{\tilde v_p^2 \tilde \tau_A^2}{\tilde\tau}\Big(e^{-\tilde \tau/\tilde\tau_A}-1+\frac{\tilde \tau}{\tilde \tau_A}\Big) \right]. 
\label{eq:q_aoup_thm_optimal}
\end{equation}
Note that the contribution of the active noise $\Big(\tilde v_p /\tilde F\Big)^2$ in the thermodynamic cost $\dot{\Sigma}^{\rm ss}(\alpha^*)$ is canceled out, but only heat dissipation due to the external driving force remains.  
Clearly, $Q(\tilde \tau) \geq Q^* (\tilde \tau) \geq 2 k_{\rm B}$ for all time and all parameters, thus, $Q^*$ provides a tighter lower bound for the MSD compared to $Q$. 
 
We demonstrate the validity of our analytic results with Langevin simulations of the single AOUP system. The results are presented in Fig.~\ref{fig3}, which shows the plots of $Q/k_{\rm B}$ and $Q^*/k_{\rm B}$ as a function of time lag $\tilde \tau$ for various P\'eclet numbers $\tilde v_p$ in the steady state. For guaranteeing the system being in a steady state, we took the data generated after a sufficiently long-time relaxation period. 
We have evaluated each marker point by averaging the displacement, MSD (variance), and $\eta \dot{x}_0$ over time and an ensemble of $100$ simulated trajectories. Further explanations of the time and ensemble averages are provided in Supplementary Section~\ref{secA:average}.
In the figure, the colored markers represent the simulated data of $Q$ (empty) and $Q^*$ (filled), while the dashed and solid lines denote the theoretical predictions for $Q$~\eqref{eq:q_aoup_thm} and $Q^*$~\eqref{eq:q_aoup_thm_optimal}, respectively. Throughout the entire time domain, we observe a notable agreement between the simulated data and the theoretical predictions for all P\'eclet numbers.

In the regime of very short time lags ($\tilde \tau \lesssim 10^{-4} \ll \tilde \tau_A$), $Q$ and $Q^*$ exhibit plateaus across all $\tilde v_p$ values. This behavior arises as the strength of the Gaussian white noise is the order of $\sqrt{\Delta t}$, while those of other factors are the order of $\Delta t$ for short-time duration $\Delta t$. Thus, the Gaussian white noise dominates over the others in this region, which leads to the linear increase of the MSD in time. As a result, $Q$ and $Q^*$ remain constant in the short-time regime. 
In the regime of sufficiently long time lags ($ \tilde \tau_A \ll 1 \lesssim \tilde \tau$), where $\tilde \tau_A$ is negligibly small compared to the time lag of a process, the active noise simply strengthens the diffusivity by an amount of $v_p^2 \tau_A$. Consequently, the MSD increases linearly in time with this increased diffusivity, which causes another plateau in this regime. 
On the other hand, in the intermediate time domain ($10^{-4} \ll \tilde \tau \lesssim \tau_A$), the self-propelled motion of the AOUP dominates the dynamics. Hence, the particle exhibits a ballistic motion, characterized by the MSD proportional to $\tilde \tau^2$. As a consequence, $Q$ and $Q^*$ increase with time in this region. 

In the absence of active noise ($\tilde v_p =0$), $Q$ and $Q^*$ collapse into the lowest TUR bound $2k_{\rm B}$. Therefore, the MSD bound calculated from the TUR always provides a tight estimation of the MSD. However, as the P\'eclet number increases, $Q$ also increases and deviates from the TUR bound over the entire time domain due to the factor $1+\tilde v_p^2 /\tilde F^2$ in Eq.~\eqref{eq:q_aoup_thm}. Therefore, the bound of the MSD from the TUR gets looser for larger $\tilde v_p$. However, such an entire increment by the factor $1+\tilde v_p^2/\tilde F^2$ is not present for $Q^*$, as seen in Eq.~\eqref{eq:q_aoup_thm_optimal}. As a result, $Q^*$ provides a tight bound of the MSD at least for a very short time regime, as illustrated in Fig.~\ref{fig3}. Nevertheless, the MSD bound calculated from $Q^*$ still gets looser in the intermediate and large time domain for a finite $\tilde v_p$. This is due to the ballistic motion of the AOUP. This implies that active noise hinders the accurate estimation of MSD.

\subsection{Case 2: An AOUP connected to a Rouse chain at its center} \label{sec:Rouse}

\begin{figure}
\centering
\includegraphics[width=8cm]{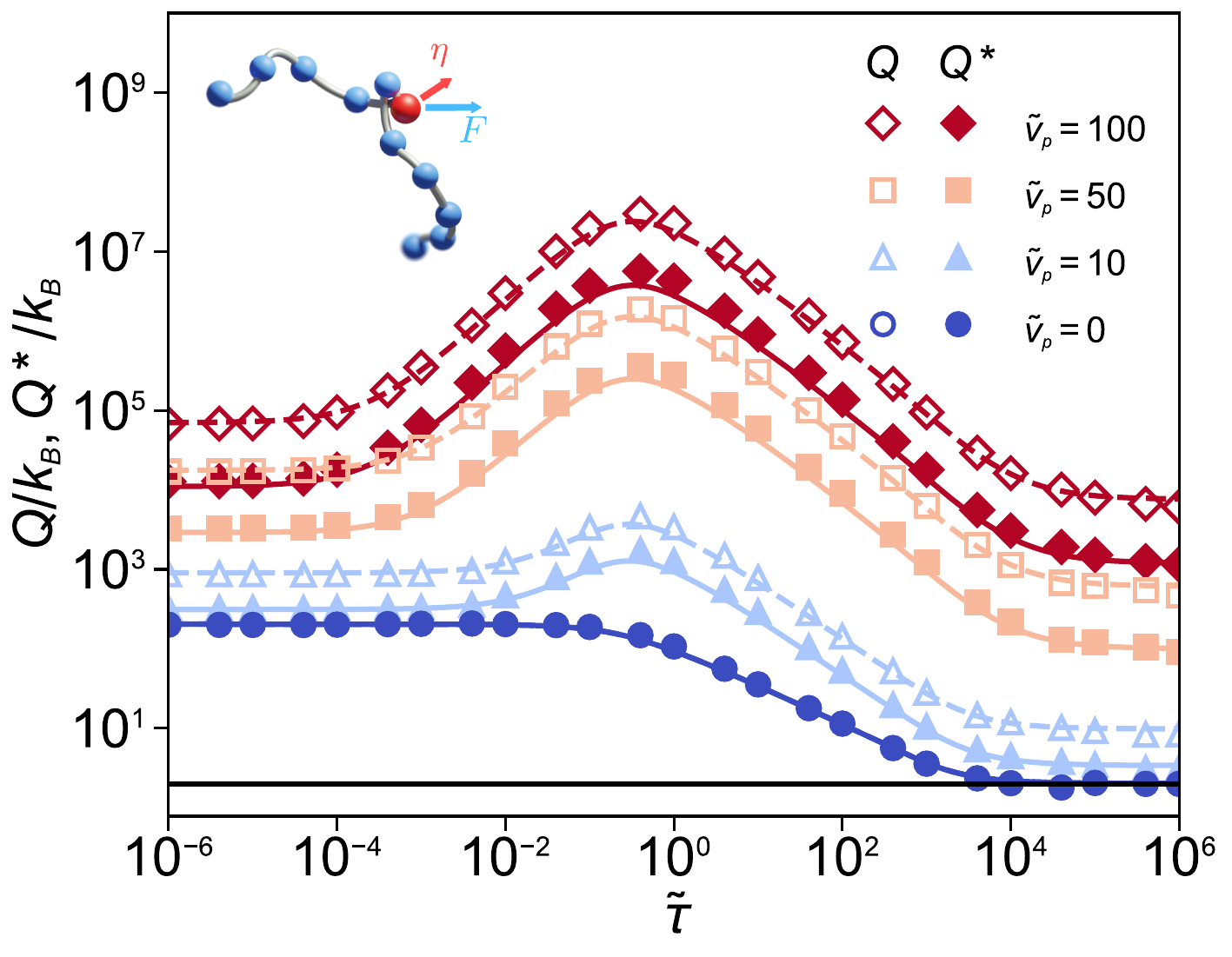}
\caption{
Plots of $Q/k_{\rm B}$ and $Q^*/k_{\rm B}$ as a function of $\tilde \tau$ for an active Rouse chain having an AOUP at its center. Empty and filled markers denote $Q/k_{\rm B}$ and $Q^*/k_{\rm B}$, respectively, for various P\'eclet numbers $\tilde v_p = 0, 10, 50$, and $100$ with a given $\tilde \tau_A=0.1$. The colored dashed and solid curves denote the analytic expressions of $Q/k_B$ and $Q^*/k_B$, respectively. The black horizontal solid line illustrates the lower limit of the TUR, i.e., $2$. 
}
\label{fig4}
\end{figure}

The second example is the one-dimensional Rouse chain where the interaction potential is given by $U^{\rm int}({\bm x}) = \frac{k}{2}\sum_{i=-M}^{M-1}(x_{i+1}-x_{i})^2$ with a spring constant $k$. The center monomer of the Rouse chain, indicating the particle indexed by $0$, is an AOUP and driven by a constant force $F$ (Fig.~\ref{fig4}, inset). 

For obtaining the analytic expressions of $Q$ and $Q^*$, it is necessary to evaluate the variance of the AOUP's displacement ${\rm Var}[\Delta \tilde x_0(\tilde \tau)]$ and the energy consumption by the active noise $\langle \tilde\eta \circ \dot{\tilde x}_0\rangle^{\rm ss}$ in the steady state. First, we recently derived the exact solution on the variance for this system in a recent study~\cite{joo2020anomalous}, the result of which is presented in Eq.~\eqref{eqA:variance_Rouse}.
This solution indicates that the presence of active noise results in a higher variance compared to its absence. Second, the analytic expression of $\langle \tilde\eta \circ \dot{\tilde x}_0\rangle^{\rm ss}$ is 
\begin{equation} \label{eq:eta_dotx_expression}
    \langle \tilde\eta \circ \dot{ \tilde x}_0 \rangle^{\rm ss} =
    \frac{ \tilde v_p^2}{2M+1}\Bigg(1+2\sum_{m=1}^{2M}\frac{\cos^2{\Big(\frac{m\pi}{2}\Big)}}{ k_m\tilde\tau_A+1}\Bigg),
\end{equation} 
where $k_m \equiv 4\tilde k \sin^2(\frac{m\pi}{4M+2})$. 
See Supplementary Section~\ref{sec:work_rouse} for the detailed derivation of Eq.~\eqref{eq:eta_dotx_expression}.  This expression shows that the energy consumption by the active noise is directly proportional to the square of the P\'eclet number and, thus, always positive across all parameter values. As a result, the presence of active noise leads to a higher value of $Q$ compared to the case without it.

Figure~\ref{fig4} shows the simulation plots of $Q$ (empty markers) and $Q^*$ (filled markers) as a function of $\tilde \tau$ along with their corresponding analytical predictions denoted by the solid and dashed lines, respectively. The simulation results are in excellent agreement with the theoretical predictions.
In the very short time domain $\tilde \tau \lesssim 10^{-4} \ll \tilde \tau_{\rm A}$, the AOUP exhibits the normal diffusive motion as the thermal noise dominates over other influences. Therefore, during this time, the MSD is linearly proportional to time, and $Q$ and $Q^*$ show no dependence on time.
Afterward, in the time lag of $10^{-4} \ll \tilde \tau \lesssim \tilde \tau_{\rm A}$, the particle exhibits superdiffusive behavoir, attributed to its self-propulsive motion induced by the active noise. This persists until the influences from the interactions with the remaining Rouse chain becomes significant. 
In the next time regime $\tilde \tau_{\rm A} \ll \tilde \tau \ll 10^4$, the collective motion of the Rouse chain leads to a subdiffusive motion, where the variance scales as $\text{Var}[\Delta x_0(\tilde \tau)]\sim \tilde \tau^ \kappa$ ($0< \kappa<1$). This results in the decline of the values $Q$ and $Q^*$ in this time domain. 
Finally, in the sufficiently long time regime $\tilde \tau \gtrsim 10^4$, the impact of the collective motion of the Rouse chain interaction diminishes. Instead, the MSD of the AOUP approaches that of the center-of-mass movement, indicating a shift toward the normal diffusion. Consequently, both $Q$ and $Q^*$ show time-independent behavior again.

\begin{figure}
\centering
\includegraphics[width=8cm]{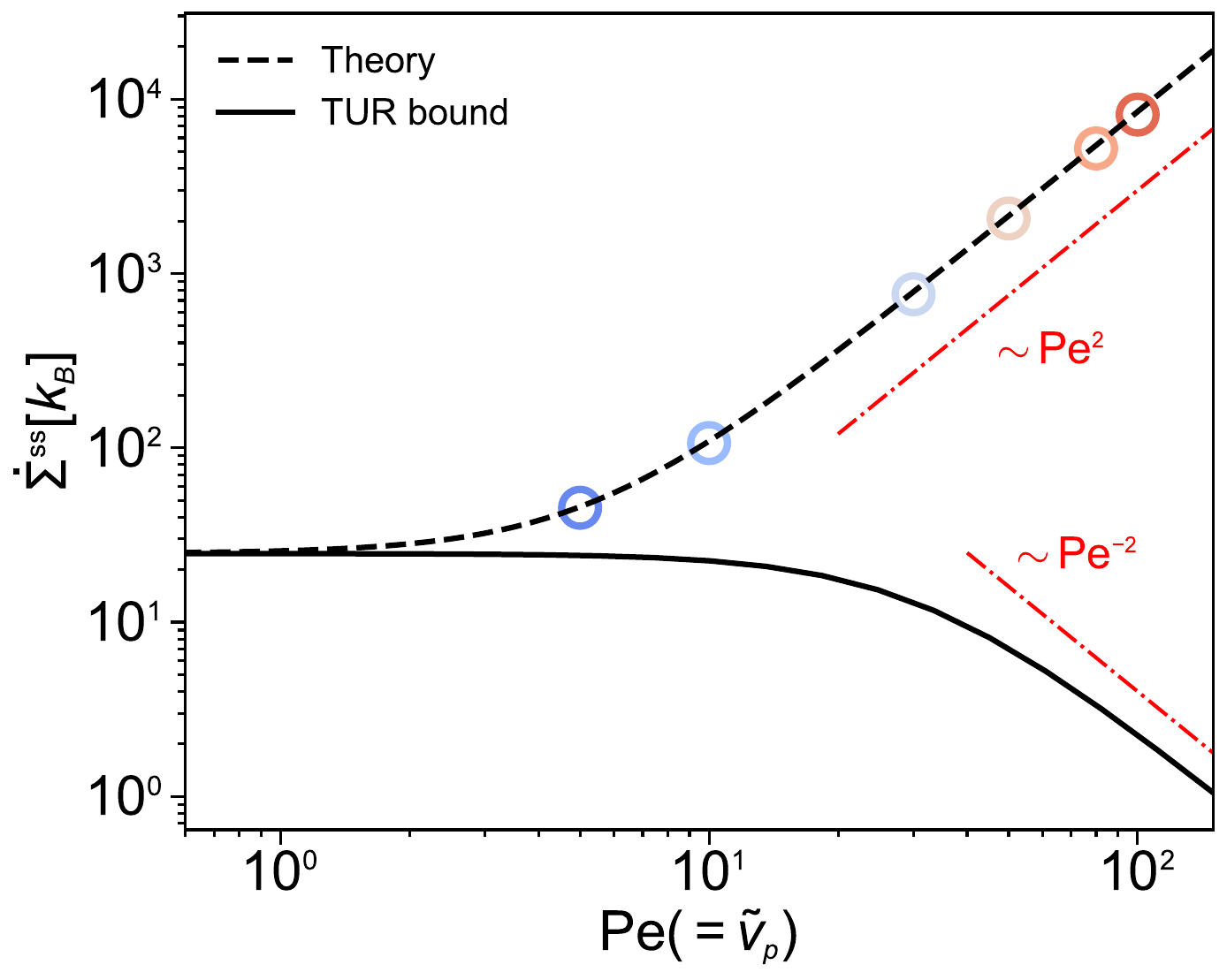}
\caption{
The rate of thermodynamic cost against Pe for the active Rouse chain shown in Fig.~\ref{fig4}. The symbols mark the numerical values from the simulation after the relaxation time ($\tilde\tau \gtrsim 10^4$) while the dashed line indicates the theoretical value of $\tilde F\langle \dot{\tilde{x}}_0\rangle  +\langle \tilde\eta \circ \dot{\tilde{x}}_0 \rangle$.
The solid line represents the TUR bound for the total dissipation rate from Eq.~\eqref{eq:dissipation_inequality}.
} 
\label{fig5}
\end{figure}

In the absence of the active noise ($\tilde v_p =0$), $Q$ and $Q^*$ collapse into a single curve, which corresponds to the result of Hartich \textit{et al.}~\cite{hartich2021thermodynamic}. In this case, the variance finally touches the lower bound $2k_{\rm B}$ at which it undergoes a transition from a subdiffusive to a normal diffusion. From this, we can accurately estimate the extent of the anomalous diffusion. 
When $\tilde{v}_p\neq 0$, both $Q$ and $Q^*$ are elevated compared to the case of $\tilde v_p =0$. This increase mainly arises because the thermodynamic cost is elevated by the energy consumption due to the active noise, which amounts to $[1-\frac{\langle \eta \circ \dot{x}_0\rangle^{\rm ss}}{\gamma v_p^2}]\langle \eta \circ \dot{x}_0\rangle^{\rm ss}>0$. In contrast to the single AOUP system, a distinct gap exists between the TUR bound and $Q^*$ for finite P\'eclet number. This hinders the precise estimation of the extent of anomalous diffusion using TUR when active noise is involved in the dynamics.

The TUR~\eqref{eq:contracted_ssTUR} can be expressed to state the lower bound for the steady-state thermodynamic cost as follows: 
\begin{equation}
\label{eq:dissipation_inequality}
     \dot{\Sigma}^{\rm ss} \geq 2k_{\rm B} \frac{\tilde F^2N^{-2} \tilde\tau}{{\rm Var}[\Delta \tilde x_0(\tilde\tau)]}\equiv \dot{\Sigma}^\mathrm{ss}_\mathrm{TUR}.
\end{equation}
In Fig.~\ref{fig5}, we plot the simulated thermodynamic cost (symbols) for the active Rouse chain system and its TUR bound, $\dot{\Sigma}^\mathrm{ss}_\mathrm{TUR}$, as a function of Péclet numbers. For comparison, we also numerically plot the theoretical expression (dashed line) for $\tilde F\langle \dot{\tilde{x}}_0\rangle  +\langle \tilde\eta \circ \dot{\tilde{x}}_0 \rangle$. While the theoretical curve agrees excellently with the simulation, the $\dot{\Sigma}^\mathrm{ss}_\mathrm{TUR}$ provides a looser bound for the thermodynamic cost as the active noise is stronger (i.e., $\dot{\Sigma}^\mathrm{ss}/\dot{\Sigma}^\mathrm{ss}_\mathrm{TUR}\sim \mathrm{Pe}^4$). 
In the large-Pe limit, the thermodynamic cost asymptotically scales as $\mathrm{Pe}^2$, because the term $\langle \tilde\eta\circ \dot{\tilde x}_0 \rangle^\mathrm{ss}$ dominates over $Fv^\mathrm{ss}$ in the second line of Eq.~\eqref{eq:Fisher_MSD_contracted} and $\langle \tilde\eta \circ \dot{\tilde x}_0 \rangle^\mathrm{ss}$ is proportional to $\mathrm{Pe}^2$ [Eq.~\eqref{eq:eta_dotx_expression}]. Meanwhile, the corresponding TUR bound exhibits the power scaling of $\sim\mathrm{Pe}^{-2}$ in the large-Pe limit, since the MSD shows Fickian motion with the effective diffusivity as $\mathrm{Var}[\Delta\tilde{x}_0(\tilde\tau)]\simeq [2+2\mathrm{Pe}^2\tilde{\tau}_A]N^{-1}\tilde\tau$.

We infer that the observed loose bound is likely a general feature for other active systems beyond AOUPs, based on the inequality relation~\eqref{eq:dissipation_inequality}. Active noise increases energy dissipation ($\dot{\Sigma}^{\rm ss}$) compared to passive systems. However, it also reduces precision by increasing the variance relative to the mean displacement, causing the thermodynamic uncertainty bound ($\dot{\Sigma}^\mathrm{ss}_\mathrm{TUR}$) to decrease as the strength of the active noise grows.

\section{Conclusions}
\label{sec:conclusion}

\begin{table}
\centering
\begin{tabular}{l c c}
\hline\hline
$\alpha$    &  parity & $T\dot\Sigma^\mathrm{ss}(\alpha)$\\\hline
0    & even & $\langle   [ -\partial_{x_0}U^{\rm int}+F+\eta]\circ\dot{x}_0 \rangle^{\rm ss}$\\
1    & odd & $\langle [-\partial_{x_0}U^{\rm int}+F] \circ [\dot{x}_0 - \gamma^{-1} \eta] \rangle^{\rm ss}$\\ 
$\alpha^*$    & mixture & $Fv^\mathrm{ss}+\langle \eta\circ\dot x_0\rangle^\mathrm{ss}  - \frac{ {\langle \eta\circ \dot{x}_0 \rangle^{\rm ss}}^2 }{\gamma v_p^2}$\\
\hline\hline\end{tabular}
\caption{Time-reversal symmetry properties of the active force depending on $\alpha$ and the corresponding thermodynamic cost.}
\label{table:active_noise_parity}
\end{table}

Using the Cram\'er--Rao inequality, we have rigorously derived the TUR for AOUP systems subject to arbitrary time-dependent protocols. A key finding is that the presence of active noise fundamentally modifies the thermodynamic cost in the TUR. As described in Eqs.~\eqref{eq:therdmodynamiccost}, \eqref{eq:Fisher_MSD_contracted}, \& \eqref{eq:tighterQ}, the modified thermodynamic cost now incorporates both the conventional entropy production associated with passive stochastic systems, as well as the energy consumption attributable to the active noise inherent in AOUP dynamics.

We applied this formalism to evaluate the lower bound of the MSD for active systems subjected to a constant force in free space. Since this system lacks a steady state, we introduced the concept of a contracted pdf. This contracted pdf is defined within a finite region of space while still obeying the same Fokker-Planck equation as the original system in free space. This mathematical manipulation allowed us to derive the steady-state TUR, even in the absence of a true steady state for the original system.
Furthermore, by introducing an optimization factor $\alpha$ into the equation of motion, we derived the {\color{black}optimized} TUR bound for the displacement of AOUP systems, as captured in Eq.~\eqref{eq:tighterQ}. 

It is important to note that the optimization factor $\alpha$ carries a significant physical interpretation. Specifically, $\alpha=0$ corresponds to an  active force with even parity under time reversal, whereas $\alpha=1$ represents an active force with odd parity. These distinctions are summarized in Table~\ref{table:active_noise_parity}. The parity of the active force has been discussed in recent works \cite{shankar2018hidden,dabelow2019irreversibility,crosato2019irreversibility,oh2023effects}. 
For a single AOUP system, the EP rates ($\dot{\Sigma}^{\rm ss}(\alpha)$) for the first two rows of Table~\ref{table:active_noise_parity} are identical with those by Shankar \textit{et al.}~\cite{shankar2018hidden}, who calculated the entropy production for active particles with both even and odd parity forces, based on energy conservation in the underdamped limit, and then took the overdamped limit.

Dabelow \textit{et al.}~\cite{dabelow2019irreversibility} provided a physical interpretation of the entropy productions following the framework of Sekimoto~\cite{sekimoto1998langevin}. For even parity (the first row in Table~\ref{table:active_noise_parity}), $\eta$ is interpreted as an additional external force.  For odd parity (the second row in Table~\ref{table:active_noise_parity}), the entropy production is generated from the hydrodynamic friction given by $-\gamma \dot{x}_0(t) + \eta(t)$, arising from the velocity difference between the particle ($\dot{x}_0$) and the surrounding fluid environment ($\eta/\gamma$). Following these interpretations, our result in Eq.~\eqref{eq:Sigma_alpha} can be understood as
\begin{equation}
\dot{\Sigma}^{\rm ss}(\alpha) = \frac{1}{T} \left\langle [-\partial_{x_0}U^{\rm int}+F + (1-\alpha)\eta] \circ [\dot{x}_0 - \gamma^{-1} \alpha \eta] \right\rangle^{\rm ss},   
\end{equation} indicating that the $\Sigma(\alpha)$ is the thermodynamic cost when the $(1-\alpha)$ portion of $\eta$ has an even parity, while the other portion of $\alpha$ has an odd parity.

However, it is worth highlighting that there remain controversies over the physical realizations of each parity. For instance, while Dabelow \textit{et al.}~\cite{dabelow2019irreversibility} assumed that the active force in self-propelled particles has odd parity, Shankar \textit{et al.}~\cite{shankar2018hidden} and Crosato \textit{et al.}~\cite{crosato2019irreversibility} argued that such particles exhibit even parity. Despite these ongoing debates, our theory determines an optimal value of $\alpha$ (denoted as $\alpha^*$) between 0 and 1, which generally yields a tighter bound on the MSD, regardless of the specific parity chosen for the active force.

From numerical simulations, we have demonstrated that this optimized bound indeed leads to a tighter inequality. Our investigations into both the single AOUP system and the active Rouse chain through this extended TUR have revealed that the presence of active noise increases both fluctuations and the thermodynamic cost, which, in turn, loosens the TUR. Consequently, unlike systems without active noise~\cite{hartich2021thermodynamic}, precise estimation of the extent of the anomalous diffusion becomes harder as the strength of the active noise increases. 

We expect that our results will serve as a foundation for more systematic studies into the fluctuating dynamics of biological systems, many of which operate in environments characterized by active noise. These systems, ranging from molecular motors to cellular processes, can benefit from the insights provided by our refined understanding of TUR in active systems. {There remain multiple open questions in stochastic thermodynamics with biological systems. For instance, extending our theory to the systems driven by L{\'e}vy noises or fractional Gaussian noises, which are often used to describe the diffusion dynamics of the biological system, presents intriguing challenges. Investigating such systems would yield valuable insights, benefiting both statistical physics and biophysics.}

\section{Supplementary Material}
Supplementary material is available at PNAS Nexus online.

\section{Funding}
This work was supported by the National Research Foundation (NRF) of S. Korea, Grant No.~RS-2023-00218927  \& No.~RS-2024-00343900 (J.-H.J) and KIAS Individual Grant No. PG064902 (J.S.L.).

\section{Data availability}
The code used to generate and analyze the simulation data, along with the associated datasets, is available at \url{https://github.com/tark1998/TUR-AOUP}.

\section{Competing Interest}
Authors declare no competing interest.

\bibliography{ref.bib}

\end{document}


\title{Supplementary Material:\\
Thermodynamic uncertainty relation for systems with active Ornstein--Uhlenbeck particles}

\author{Hyeong-Tark Han}
\author{Jae Sung Lee}
\author{Jae-Hyung Jeon}

\maketitle

\subsection{Detailed derivation of TUR}\label{secA:detailed_derivation}
$P_\theta$ is written in terms of the pdf of the original dynamics $P$ in Eq.~\eqref{eq:FP} as
\begin{equation} \label{eq:PandP_theta}
    P_\theta (\bm{x},\bm{\eta},t;\omega) = P (\bm{x},\bm{\eta},t_\theta;\omega_\theta),
\end{equation}
where $t_\theta \equiv (1+\theta) t$ and $\omega_\theta \equiv \omega/(1+\theta)$. We can demonstrate the relation of  Eq.~\eqref{eq:PandP_theta} by directly plugging it into  Eq.~\eqref{eq:perturbedFP}. This perturbation Eq.~\eqref{eq:perturbedFP} amounts to applying an additional force to the original Langevin dynamics Eqs.~\eqref{eq:eom_aoup} and \eqref{eq:AOUPeq} as follows:
\begin{align}
    &\gamma \dot{x}_i = f_i (\bm{x},\bm{\eta},\omega t)+\theta \gamma \frac{J_{x_i,\theta}}{P_\theta} +\eta_i+ \xi_i, \nonumber \\
     & \tau_{\rm A}\dot{\eta}_i = -\eta_i +\theta \tau_{\rm A} \frac{J_{\eta_i,\theta}}{P_\theta} + \zeta_i,
    \label{eq:eom_aoup_perturb}
\end{align}
where $J_{x_i,\theta}$ and $J_{\eta_i,\theta}$ are defined as
\begin{align}
    J_{x_i,\theta}  \equiv  \left(\frac{g_i}{\gamma} - \frac{k_BT}{\gamma} \partial_{x_i} \right) P_\theta ,~~~~
    J_{\eta_i,\theta}  \equiv -\left(\frac{\eta_i}{\tau_A}+\frac{\gamma^2v_p^2}{\tau_A} \partial_{x_i} \right) P_\theta
\end{align}
{with $g_i = g_i ({\bm{x}},\bm{\eta},t;\omega) \equiv f_i +\eta_i$.}
Similarly to the pdf above, $J_{x_i,\theta}$ and $J_{\eta_i,\theta}$ are related to $J_{x_i}$ and $J_{\eta_i}$ as
\begin{align}
    &J_{x_i,\theta} (\bm{x},\bm{\eta},\omega; t) = J_{x_i} (\bm{x},\bm{\eta},\omega_\theta ; t_\theta), \nonumber \\
    &J_{\eta_i,\theta} (\bm{x},\bm{\eta},\omega; t) = J_{\eta_i} (\bm{x},\bm{\eta},\omega_\theta; t_\theta).
\end{align}
Note that the probability currents of the perturbed system are given by multiplying $J_{x_i,\theta}$ and $J_{\eta_i,\theta}$ by $(1+\theta)$. Within this perturbed dynamics, the mean value of the observable Eq.~\eqref{eq:obs_over} is evaluated as 
\begin{equation} \label{eq:obs_pertur}
    \langle \Theta_\tau(\omega) \rangle_\theta = \int_0^\tau dt\int d{\bm x}d{\bm \eta} \; {\bm\Lambda}({\bm x}, {\bm \eta}, \omega t)^{\textsf T} (1+\theta){\bm J}_{{\bm x},\theta} (\bm{x},\bm{\eta},\omega; t),
\end{equation}
where $\langle \cdots \rangle_\theta$ represents the average in the perturbed dynamics parameterized by $\theta$ and ${\bm J}_{{\bm x},\theta} = (J_{x_1,\theta}, \cdots, J_{x_N,\theta})^{\textsf T}$. Equation~\eqref{eq:obs_avg} is used for deriving  Eq.~\eqref{eq:obs_pertur}. 

First, calculation of $\partial_\theta \langle \Theta_\tau(\omega) \rangle_\theta |_{\theta=0 }$ is straightforward from  Eq.~\eqref{eq:obs_pertur}. By changing the variables $t$ and $\omega$ into $t_\theta$ and $\omega_\theta$, we have
\begin{align} \label{eq:obs_partial}
    &\partial_\theta \langle \Theta_\tau(\omega) \rangle_\theta |_{\theta=0} \nonumber \\
    &= \left. \partial_\theta \int_0^{\tau_\theta} dt_\theta \int d{\bm x}d{\bm \eta} \; {\bm\Lambda}({\bm x}, {\bm \eta}, \omega_\theta t_\theta)^{\textsf T} {\bm J}_{{\bm x}} ({\bm x}, {\bm \eta}, \omega_\theta ; t_\theta) \right|_{\theta=0}  \nonumber \\
    &= \partial_\theta \langle \Theta_{\tau_\theta} (\omega_\theta)\rangle |_{\theta=0} =  \hat{h}\langle \Theta_{\tau} (\omega)\rangle, 
\end{align}
where $\tau_\theta \equiv (1+\theta) \tau$ and $\hat{h} \equiv \tau\partial_\tau - \omega \partial_\omega$.


Next, we calculate the Fisher information of path probability. We consider a stochastic trajectory $\Gamma$ generated by the perturbed overdamped Langevin equation~(\ref{eq:eom_aoup_perturb}) during the time $0\leq t\leq\tau$. From the Onsager-Machlup formalism~\cite{onsager1953fluctuations}, the probability $\mathcal{P}_\theta (\Gamma)$ for observing the trajectory $\Gamma$ is
\begin{equation}
    \mathcal{P}_\theta (\Gamma) = \mathcal{N} P^{\rm int} \exp\left[ -\int_0^{\tau} dt \mathcal{A}\right], 
\end{equation}
where $\mathcal{N}$ is a normalization factor, $P^{\rm int} \equiv P ({\bm x}, {\bm \eta}, 0)$ represents the initial pdf, and the action functional $\mathcal{A}$ is given by
\begin{align}
    \mathcal{A}
    =&\sum_i \frac{
    \left( \gamma \dot{x}_i - g_i -\theta \gamma \frac{J_{x_i,\theta}}{P_\theta}\right)^2}{4\gamma k_BT} 
    +\sum_i \frac{\left( \tau_A\dot{\eta}_i  +\eta_i -\theta \tau_A \frac{J_{\eta_i , \theta} }{P_\theta} \right)^2}{4\tau_A\gamma^2v_p^2 }.
  \label{eq:theta-trajectory}
\end{align}
Note that $\mathcal N$ and $P^{\rm init}$ are independent of $\theta$. Therefore, we can estimate the Fisher information at $\theta =0$ as
\begin{align} \label{eqA:Fisher_0_int}
    \mathcal{I} (0) &= \langle -\partial_\theta^2 \ln \mathcal{P}_\theta (\Gamma) \rangle_\theta |_{\theta=0}  = \left.  -\left\langle \int_0^\tau dt \partial_\theta^2 \mathcal{A}  \right\rangle_\theta \right|_{\theta=0}. 
\end{align}
We can manipulate the integrand of Eq.~(\ref{eqA:Fisher_0_int}) as
\begin{align} \label{eqA:integrandA}
    \langle \partial_\theta^2 \mathcal{A} \rangle |_{\theta = 0} 
    &=  \sum_i  \left. \left\langle \partial_\theta^2  \frac{(\gamma \dot{x}_i - g_i)^2 -2\theta \gamma \frac{J_{x_i,\theta}}{P_\theta} \bullet \xi_i   -\theta^2 \gamma^2 \frac{J_{x_i,\theta}^2}{P_\theta^2} }{4\gamma k_{\rm B} T}  \right\rangle \right|_{\theta=0} 
    + \sum_i  \left. \left\langle \partial_\theta^2 \frac{ (\tau_{\rm A} \dot{\eta}_i + \eta_i )^2 -2\theta \tau_{\rm A} \frac{J_{\eta_i,\theta}}{P_\theta} \bullet \zeta_i   -\theta^2 \tau_{\rm A}^2 \frac{J_{\eta_i,\theta}^2}{P_\theta^2}}{4 \tau_{\rm A} \gamma^2 v_p^2 } \right\rangle \right|_{\theta=0} \nonumber \\
    &=- \sum_i \left\langle \frac{\gamma J_{x_i}^2}{2k_{\rm B} T P^2} + \frac{\tau_{\rm A} J_{\eta_i}^2}{2 \gamma^2 v_p^2  P^2} \right\rangle = - \sum_i  \int d{\bm x} d{\bm \eta} \left( \frac{\gamma J_{x_i}^2}{2k_{\rm B} T P} + \frac{\tau_{\rm A} J_{\eta_i}^2}{2 \gamma^2 v_p^2  P} \right), 
\end{align}
where $\bullet$ denotes the It\^o product. For the second equality of Eq.~(\ref{eqA:integrandA}), $\langle \cdots \bullet \xi_i \rangle = \langle \cdots \rangle \langle \xi_i \rangle =0$ and $\langle \cdots \bullet \eta_i \rangle = \langle \cdots \rangle \langle \eta_i \rangle =0$ are used. Therefore, the Fisher information at $\theta = 0$ becomes 
\begin{align} \label{eqA:Fisher_0}
    \mathcal{I} (0) =  \frac{1}{2k_{\rm B}} \int_0^\tau dt \int d{\bm x} d{\bm \eta} \sum_i \left( \frac{ \gamma J_{x_i}^2}{ T P} +\frac{k_B\tau_A J_{\eta_i}^2}{\gamma^2v_p^2P} \right).
\end{align}
Note that the first term on the right-hand side of  Eq.~\eqref{eqA:Fisher_0} is identical to the conventional EP, while the second term originates from the presence of the active noise. Let the Fisher information at $\theta=0$ as 
\begin{equation} \label{eq:Fisher1}
    \mathcal{I}(0) = \frac{1}{2k_{\rm B}} \Sigma_\tau , 
\end{equation}
where $\Sigma_\tau$ represents the thermodynamic cost. Hence, the thermodynamic cost $\Sigma_\tau=2k_\mathrm{B}\mathcal{I}(0)$ is always nonnegative by definition.

Substituting the Cram\'er--Rao inequality Eq.~\eqref{eq:CramerRao} with  Eq.~\eqref{eq:obs_partial}, and \eqref{eq:Fisher1} finally leads to the  TUR:
\begin{equation}
    \frac{{\rm Var}[\Theta_\tau (\omega)]}{[\hat{h}\langle \Theta_\tau (\omega) \rangle]^2} \Sigma_\tau \geq 2 k_{\rm B}. 
\end{equation}

\subsection{Physical insight on $\Sigma_\tau$}\label{secA:sigma_insight}
To gain a physical insight on $\Sigma_\tau$, we need to identify the physical ingredients contributing to it. For this purpose, we can write the rate of the system entropy change as
\begin{align} \label{eq:sysEPrate}
    -k_{\rm B}\frac{d}{dt}\ln P = -k_{\rm B} \sum_i \left( \partial_{x_i} \ln P \circ \dot{x}_i + \partial_{\eta_i} \ln P \circ \dot{\eta}_i \right) -k_{\rm B}\partial_t \ln P .
\end{align}
Using the relations
\begin{align}
    \frac{\partial_{x_i} P}{P} = \frac{\gamma}{k_{\rm B}T} \left( \frac{g_i}{\gamma} - \frac{J_{x_i}}{P} \right), ~~\frac{\partial_{\eta_i} P}{P} = \frac{\tau_{\rm A}}{\gamma^2 v_p^2} \left(- \frac{\eta_i}{\tau_{\rm A}} - \frac{J_{\eta_i}}{P} \right)
\end{align}
from Eq.~\eqref{eq:prob_curr_def}, we can manipulate  Eq.~\eqref{eq:sysEPrate} as
\begin{align}
    &-k_{\rm B}\frac{d}{dt}\ln P +\sum_i \left( \frac{1}{T} f_i \circ \dot{x}_i + \frac{1}{T}  \eta_i \circ \dot{x}_i - \frac{k_{\rm B}}{\gamma^2 v_p^2 }\eta_i \circ \dot{\eta}_i \right) \nonumber \\
    & = -k_{\rm B} \partial_t \ln P + \sum_i \left( \frac{\gamma J_{x_i}}{TP} \circ \dot{x}_i + \frac{k_{\rm B} \tau_{\rm A} J_{\eta_i}}{\gamma^2 v_p^2 P} \circ \dot{\eta}_i\right) \equiv \sigma.
    \label{eq:sigma_expression} 
\end{align}
Note that, in  Eq.~\eqref{eq:sigma_expression}, $-k_{\rm B} \ln P = S_{\rm sys}$ denotes the system entropy and $- f_i \circ \dot{x}_i = \dot{Q}_i^{f}$ represents the conventional heat rate injected into $i$th particle. Averaging $\sigma$ yields 
\begin{align} \label{eq:sigma_avg}
    \langle \sigma \rangle & = \langle \dot{S}_{\rm sys} \rangle -\sum_i  \frac{\langle \dot{Q}_i^f\rangle}{T} + \sum_i \frac{\langle \eta \circ \dot{x}_i \rangle}{T}   \nonumber \\
    & =\int d{\bm x} d{\bm \eta} \sum_i \left( \frac{ \gamma J_{x_i}^2}{ T P} +\frac{k_B\tau_A J_{\eta_i}^2}{\gamma^2v_p^2P} \right) = \frac{d}{dt}\Sigma_t.
\end{align}
For deriving  Eq.~\eqref{eq:sigma_avg}, $\frac{d}{dt} \langle \eta_i^2 \rangle = 2\langle \eta_i \circ \dot{\eta}_i \rangle = 0 $ in the steady state of $\eta_i$ and $\langle \partial_t \ln P \rangle =0$ are used. 

\subsection{Fisher information for the contracted pdf}\label{secA:FisherInformation_contractedpdf}
Similarly to  Eq.~\eqref{eq:eom_aoup_perturb}, the perturbed Fokker--Planck equation~\eqref{eq:perturbed_MSD_FP} in the steady state corresponds to the following Langevin dynamics
\begin{align}
    &\gamma \dot{x}_i = -\partial_{x_i} U^{\rm int} ({\bm x}) +\delta_{i,0} ( F+\eta) +\theta \gamma \frac{\hat{J}_{x_i} \bar{P}^{\rm ss}}{\bar{P}^{\rm ss}} + \xi_i, \nonumber \\
     & \tau_{\rm A}\dot{\eta} = -\eta +\theta \tau_{\rm A} \frac{\hat{J}_{\eta}\bar{P}^{\rm ss}}{\bar{P}^{\rm ss}} + \zeta,
    \label{eq:eom_aoup_contracted_perturb}
\end{align}
with a shifted origin. Following the similar calculation performed in Supplementary Section~\ref{secA:detailed_derivation}, we can obtain the Fisher information at $\theta =0$ as
\begin{align} \label{eq:Fisher_contracted}
    \mathcal{I}(0) &= \frac{1}{2k_{\rm B}}  \int_0^\tau dt \int d{\bm x} d\eta \left[ \sum_i  \frac{ \gamma (\hat{J}_{x_i}\bar{P}^{\rm ss})^2}{ T \bar{P}^{\rm ss}} +\frac{k_B\tau_A (\hat{J}_{\eta}\bar{P}^{\rm ss})^2}{\gamma^2v_p^2 \bar{P}^{\rm ss}} \right]. 
\end{align}
Furthermore, following the steps outlined from Eqs.~\eqref{eq:sysEPrate} to \eqref{eq:sigma_avg}, we arrive at
\begin{align}
\mathcal{I}(0) &=  \int_0^\tau dt \frac{ \sum_i  \langle -\partial_{x_i} U^{\rm int}({\bm x}) \circ \dot{x}_i \rangle^{\rm ss} + F \langle \dot{x}_0 \rangle^{\rm ss} + \langle \eta \circ  \dot{x}_0 \rangle^{\rm ss} }{2k_{\rm B}T} \nonumber \\
&= \frac{1}{2k_{\rm B}} \left(\frac{F v^{\rm ss} \tau}{ T} + \frac{ \langle \eta  \circ \dot{x}_0 \rangle^{\rm ss} \tau}{T}\right) = \frac{\dot{\Sigma}^{\rm ss}\tau}{2 k_{\rm B}},
\end{align}
where $\dot{\Sigma}^{\rm ss} = (Fv^{\rm ss} + \langle \eta \circ \dot{x}_0\rangle^{\rm ss})/T$. For the second equality, $\langle \partial_{x_i} U^{\rm int}({\bm x}) \circ \dot{x}_i \rangle^{\rm ss} = \frac{d}{dt}  \langle U^{\rm int} ({\bm x})\rangle^{\rm ss} =0$ is used.

\subsection{Tighter bound for MSD}
\label{sec:tighter_bound}

The Fokker--Planck equation for the pdf $P({\bm x},\eta,t;h) = P(x_0, x_{\pm 1}, \cdots, x_{\pm M},\eta,t;h)$ corresponding to the Langevin equation~(\ref{eq:MSDeom_h}) can be expressed as 
\begin{equation} \label{eqA:FP_tighter}
    \partial_t P = -\sum_{i} (\partial_{x_i} \hat{J}_{x_i}^{(1)} + \partial_{x_i} \hat{J}_{x_i}^{(2)})P - \partial_\eta \hat{J}_\eta P,
\end{equation}
where $\hat{J}_{x_i}^{(1)} \equiv \delta_{i,0} h\alpha \eta/\gamma$, $\hat{J}_{x_i}^{(2)} \equiv [-\partial_{x_i} U^{\rm int} +\delta_{i,0} \{F+(1-\alpha)\eta \} - k_{\rm B}T\partial_{x_i} ]/\gamma$, and $\hat{J}_\eta \equiv (-\eta-\gamma^2 v_p^2 \partial_\eta)/\tau_{\rm A}$. The same technique, dealing with the contracted pdf introduced in Sec.~\ref{sec:contractedPDF}, can be used to obtain the steady-state TUR of the Fokker--Planck equation~(\ref{eqA:FP_tighter}). To apply the Cram\'er--Rao inequality, we consider the following $\theta$-perturbed Fokker--Planck equation for the contracted pdf  in the shifted coordinate:
\begin{equation} \label{eqA:FP_pert_tighter}
    \partial_t \bar{P}_\theta = -\sum_{i} \partial_{x_i} \hat{J}_{x_i}^{(1)}\bar{P}_\theta - (1+\theta) \left[\sum_{i}   \partial_{x_i} \hat{J}_{x_i}^{(2)}   + \partial_\eta \hat{J}_\eta \right] \bar{P}_\theta,
\end{equation}
where the perturbation factor $1+\theta$ is multiplied to $\hat{J}_{x_i}^{(2)}$ and $\hat{J}_\eta$, but not $\hat{J}_{x_i}^{(1)}$.  It is straightforward to see that $\bar{P}_\theta ({\bm x},\eta,t;h) = \bar{P} ({\bm x},\eta,t_\theta;h_\theta)$, where $t_\theta \equiv (1+\theta)t$ and $h_\theta \equiv h/(1+\theta)$. Then, the steady-state mean value of the observable $\Theta_\tau(h) = \int_0^\tau dt\; \dot{x}_0 $, i.e., the displacement of the particle with index $0$, in the perturbed dynamics is given by
\begin{align} \label{eqA:mean_obs_tighter}
    \langle \Theta_\tau (h) \rangle_\theta^{\rm ss} & = \int_0^\tau dt\int d{\bm x}d\eta \left[\hat{J}_{x_0}^{(1)} + (1+ \theta) \hat{J}_{x_0}^{(2)} \right] \bar{P}_\theta^{\rm ss} ({\bm x},\eta,t;h) \nonumber \\
    & = \int_0^{\tau_\theta} dt^\prime \int d{\bm x}d\eta \left[ \frac{h_\theta\alpha \eta}{\gamma} +  \hat{J}_{x_0}^{(2)} \right] \bar{P}^{\rm ss} ({\bm x},\eta,t^\prime;h_\theta) \nonumber \\
    & = \langle \Theta_{\tau_\theta} (h_\theta)\rangle^{\rm ss}.  
\end{align}
The variables $t$ and $h$ are changed to $t^\prime = (1+\theta) t $ and $h_\theta = h/(1+\theta)$ for the second equality in Eq.~(\ref{eqA:mean_obs_tighter}), respectively. As the mean velocity $\langle \dot{x}_i \rangle^{\rm ss}$ is the same as that of Eq.~(\ref{eq:MSDeom}) regardless of the value of $h$, $\langle \Theta_{\tau_\theta} (h_\theta) \rangle^{\rm ss} = v^{\rm ss} \tau_\theta $. Therefore, we have 
\begin{equation} \label{eqA:tighter_denominator}
    \partial_\theta  \langle \Theta_\tau(h)\rangle_\theta^{\rm ss} |_{\theta=0} = v^{\rm ss} \tau.
\end{equation}

The perturbed Fokker--Planck equation~(\ref{eqA:FP_pert_tighter}) corresponds to the following Langevin dynamics:
\begin{align} \label{eq:MSDeom_pertur_tighter}
    &\gamma \dot{x}_i = -\partial_{x_i} U^{\rm int}({\bm x}) + \delta_{i,0} [F+(1-\alpha + h \alpha )\eta ] +\theta \gamma \frac{\hat{J}_{x_i}^{(2)}\bar{P}_\theta}{\bar{P}_\theta}  + \xi_i, \nonumber \\
    & \tau_{\rm A} \dot{\eta} = -\eta + \theta \tau_{\rm A} \frac{\hat{J}_{\eta}\bar{P}_\theta}{\bar{P}_\theta} +\zeta,
\end{align}
with the shifted origin. Progressing through the steps presented in Appendix~\ref{secA:detailed_derivation}, we arrive at the expression for the Fisher information at $\theta = 0$ in the steady state as follows:
\begin{align} \label{eqA:Fisher_contracted_tighter}
    \mathcal{I}(0) &= \frac{1}{2k_{\rm B}}  \int_0^\tau dt \int d{\bm x} d\eta \left[ \sum_i  \frac{ \gamma (\hat{J}_{x_i}^{(2)}\bar{P}^{\rm ss})^2}{ T \bar{P}^{\rm ss}} +\frac{k_B\tau_A (\hat{J}_{\eta}\bar{P}^{\rm ss})^2}{\gamma^2v_p^2 \bar{P}^{\rm ss}} \right]. 
\end{align}
Moreover, similar to Eq.~(\ref{eq:sigma_expression}), we have
\begin{align}
    &-\sum_i \frac{\partial_{x_i} U^{\rm int}\circ \dot{x}_i}{T}  + \frac{[F+(1-\alpha )\eta ]\circ \dot{x}_0}{T}     - \frac{k_{\rm B}}{\gamma^2 v_p^2 }\eta \circ \dot{\eta}  \nonumber \\
    & =  \sum_i  \frac{\gamma \hat{J}_{x_i}^{(2)} \bar{P}^{\rm ss} }{T \bar{P}^{\rm ss} } \circ \dot{x}_i + \frac{k_{\rm B} \tau_{\rm A} \hat{J}_{\eta} \bar{P}^{\rm ss}}{\gamma^2 v_p^2 \bar{P}^{\rm ss}} \circ \dot{\eta}. \label{eqA:sigma_expression_tighter} 
\end{align}
The steady-state condition $\frac{d}{dt} \ln \bar{P}^{\rm ss} = \partial_t \bar{P}^{\rm ss} =0$ is used for obtaining Eq.~(\ref{eqA:sigma_expression_tighter}). Taking the steady-state average of Eq.~(\ref{eqA:sigma_expression_tighter}) and applying the relations $\langle \cdots \circ \dot{x}_i \rangle^{\rm ss} = \int d{\bm x} d\eta \; \cdots (\hat{J}_{x_i}^{(1)} + \hat{J}_{x_i}^{(2)}) \bar{P}^{\rm ss}$ and $\int d{\bm x}d\eta \; \eta \hat{J}_{x_0}^{(2)} \bar{P}^{\rm ss} = \langle \eta\circ (\dot{x}_0 - h\alpha\eta/\gamma) \rangle^{\rm ss} $  to it result in 
\begin{align} \label{eqA:sigma_expression_tighter1}
    &\int d{\bm x} d\eta \left[ \sum_i  \frac{ \gamma (\hat{J}_{x_i}^{(2)}\bar{P}^{\rm ss})^2}{ T \bar{P}^{\rm ss}} +\frac{k_B\tau_A (\hat{J}_{\eta}\bar{P}^{\rm ss})^2}{\gamma^2v_p^2 \bar{P}^{\rm ss}} \right] \nonumber \\
    & = \frac{F}{T}\langle \dot{x}_0\rangle^{\rm ss} + \frac{\langle \eta \circ \dot{x}_0 \rangle^{\rm ss}}{T} -\frac{(1+h)\alpha}{T} \langle \eta \circ \dot{x}_0\rangle^{\rm ss} + \frac{h^2 \alpha^2}{T\gamma} \langle \eta^2\rangle^{\rm ss} .
\end{align}
By plugging Eqs.~(\ref{eqA:tighter_denominator}), (\ref{eqA:sigma_expression_tighter}), and (\ref{eqA:sigma_expression_tighter1}) into the Cram\'er--Rao inequality and using the  relation $\langle \eta^2\rangle^{\rm ss} = \gamma^2 v_p^2$, we reach
\begin{align}
    \frac{{\rm Var}[x_0]}{{v^{\rm ss}}^2 \tau} \left[ \dot{\Sigma}^{\rm ss} + \frac{\gamma h^2 v_p^2}{T} \alpha^2 - \frac{(1+h)\alpha}{T} \langle \eta \circ \dot{x}_0 \rangle^{\rm ss} \right] \geq 2 k_{\rm B},   
\end{align}
where $ \dot{\Sigma}^{\rm ss} = (F v^{\rm ss} +\langle \eta \circ \dot{x}_0 \rangle^{\rm ss})/T$. This TUR is reduced to Eq.~(\ref{eq:tighter_TUR}) at $h=1$.

\subsection{The MSD of a single AOUP in free space}
\label{sec:MSD_singleAOUP}
From Eq.~(\ref{eq:MSDeom}) with $U^{\rm int}=0$, $N=0$, and $\tilde x_0=\tilde x$, the displacement of $\tilde x$ from time $0$ to $\tilde \tau$ is given by
\begin{equation} \label{eqA:displacementX}
\Delta \tilde{x} (\tilde{\tau})=\int_0^{\tilde{\tau}} [\tilde{F}+\tilde{\eta}+\tilde{\xi} ]d\tilde{t}.
\end{equation}
Using the relation $\langle \tilde \eta\rangle = \langle \tilde \xi \rangle=0 $, it is straightforward to show that $\langle \Delta \tilde{x}(\tilde{\tau}) \rangle=\tilde{F}\tilde{\tau}$. In addition, the mean square of Eq.~(\ref{eqA:displacementX}) leads to
\begin{align} \label{eqA:meansquare}
\langle \Delta \tilde x(\tilde\tau)^2 \rangle&=\int_0^{\tilde\tau}\int_0^{\tilde\tau} d\tilde t_1d\tilde t_2[\tilde F^2+\langle\tilde\eta(\tilde t_1)\tilde\eta(\tilde t_2)\rangle+\langle\tilde\xi(\tilde t_1)\tilde\xi(\tilde t_2)\rangle] \nonumber
\\&=\tilde F^2\tilde\tau^2 + 2\tilde\tau+2 \tilde v_p^2 \tilde\tau_A \left[ \tilde \tau -\tilde \tau_{\rm A}( 1- e^{-\tilde \tau/\tilde \tau_A}) \right]. 
\end{align}
The autocorrelation functions $\langle \tilde \eta(t) \circ \tilde \eta (t^\prime)\rangle = \tilde v_p^2 \exp(-|t-t^\prime|/\tilde \tau_{\rm A})$ and $\langle \tilde \xi(t) \tilde \xi(t^\prime)\rangle = 2\delta (t-t^\prime)$ are used to obtain Eq.~(\ref{eqA:meansquare}). Therefore, the variance of the displacement $\langle \Delta \tilde x(\tilde\tau)^2 \rangle-\langle \Delta \tilde x(\tilde\tau) \rangle^2$ is expressed by Eq.~(\ref{eq:MSD_singleAOUP}) in the main text.

\subsection{Time and trajectory-ensemble average} \label{secA:average}

We evaluate the mean values of displacement, variance, and $\eta \dot{x}$ by averaging over time and independent samples of $N_{\rm E} = 100$ simulated trajectories, where each trajectory starts at $t=0$ and ends at $\tilde t_f$. Specifically, if we define $\Delta \tilde x^{(n)}(\tilde t+\tilde \tau, \tilde t) \equiv \tilde x^{(n)}(\tilde t+ \tilde \tau) - \tilde x^{(n)}(\tilde t)$, with $\tilde x^{(n)}(\tilde t)$ denoting the position of the particle in the $n$th trajectory at time $\tilde t$, the time and ensemble average of the displacement $\langle \overline{\Delta \tilde x(\tilde \tau)} \rangle$ is evaluated as follows:
\begin{equation}
    \langle \overline{\Delta \tilde x(\tilde \tau)} \rangle = \frac{1}{N_{\rm E}} \sum_{n=1}^{N_{\rm E}} \frac{1}{\tilde t_f -\tilde \tau} \int_0^{\tilde t_f -\tilde \tau} d\tilde t \; \Delta \tilde x^{(n)}(\tilde t+\tilde \tau, \tilde t),
\end{equation}
where $\overline{\cdots}$ and $\langle\cdots\rangle$ represent the average over time and ensemble of trajectories, respectively. Similarly, the time and ensemble averages of the variance  and $\eta \dot{x}_0$ are evaluated as
\begin{align}
    &\langle \overline{ {\rm Var}[\Delta \tilde x(\tilde \tau)] } \rangle \nonumber \\ 
    &= \frac{1}{N_{\rm E}} \sum_{n=1}^{N_{\rm E}} \frac{1}{\tilde t_f -\tilde \tau} \int_0^{\tilde t_f -\tilde \tau} d\tilde t \; \left[ \Delta \tilde x^{(n)}(\tilde t+\tilde \tau, \tilde t) -\overline{\Delta \tilde x^{(n)}(\tilde \tau)} \right]^2, \nonumber \\
    &\langle \overline{\eta \dot{x} }\rangle = \frac{1}{N_{\rm E}} \sum_{n=1}^{N_{\rm E}} \frac{1}{\tilde t_f } \int_0^{\tilde t_f} d\tilde t \; \eta^{(n)}(\tilde t) \dot{x}^{(n)}(\tilde t),
\end{align} 
where $\overline{\Delta \tilde x^{(n)}(\tilde \tau)} \equiv \frac{1}{\tilde t_f -\tilde \tau} \int_0^{\tilde t_f -\tilde \tau} d\tilde t \; \Delta \tilde x^{(n)}(\tilde t+\tilde \tau, \tilde t) $.

\subsection{The displacement variance for an AOUP connected to a Rouse chain} \label{secA:analytic_expression_variance}

The paper by Joo \textit{et al.}~\cite{joo2020anomalous} provides the analytic expression of the variance $\mathrm{Var}[\Delta \tilde x_0(\tilde\tau)]$ for the motion of an AOUP linked to a Rouse chain (Sec.~\ref{sec:Rouse}). Here we recapitulate this result as follows:
\begin{equation} \label{eqA:variance_Rouse}
    \mathrm{Var}[\Delta \tilde x_0(\tilde\tau)]=H^{(1)}(\tilde\tau)+H^{(2)}(\tilde\tau)+H^{(3)}(\tilde\tau),
\end{equation}
where $H^{(1)}(\tilde\tau)$, $H^{(2)}(\tilde\tau)$, and $H^{(3)}(\tilde\tau)$ are identified to
\begin{align}
H^{(1)}(\tilde\tau)\equiv &\frac{\tilde\tau}{M+\frac{1}{2}} +\frac{\tilde v_p^2\tilde\tau_A}{2(M+\frac{1}{2})^2}\left[\tilde\tau -\tilde\tau_A \left( 1-e^{-\frac{\tilde\tau}{\tilde\tau_A}} \right) \right], \nonumber \\
H^{(2)}(\tilde\tau) \equiv &\sum_{m=1}^{2M} \frac{\tilde\tau_R\tilde v_p^2}{(M+\frac{1}{2})^2m^2}\cos^2{\left(\frac{m\pi}{2}\right)} \nonumber \\
&\times\left(\frac{2-e^{-\frac{m^2\tilde\tau}{\tilde\tau_R}}-e^{-\frac{\tilde\tau}{\tilde\tau_A}}}{\frac{1}{\tilde\tau_A} + \frac{m^2}{\tilde\tau_R}}-\frac{e^{-\frac{m^2\tilde\tau}{\tilde\tau_R}}-e^{-\frac{\tilde\tau}{\tilde\tau_A}}}{\frac{1}{\tilde\tau_A}-\frac{m^2}{\tilde\tau_R}}\right) \nonumber \\ 
&+\sum_{m=1}^{2M}\frac{2\tilde\tau_R}{(M+\frac{1}{2})m^2}\left(1-e^{-\frac{m^2\tilde\tau}{\tilde\tau_R}}\right)\cos^2\left(\frac{m\pi}{2}\right), \nonumber \\
H^{(3)}(\tilde\tau)\equiv &\sum_{m=1}^{2M}\sum_{n=1}^{2M}\frac{2\tilde\tau_R\tilde v_p^2}{(M+\frac{1}{2})^2(m^2+n^2)}\cos^2{\left(\frac{m\pi}{2}\right)}\cos^2{\left(\frac{n\pi}{2}\right)} \nonumber\\
&\times\left[\frac{1-e^{-\frac{m^2\tilde\tau}{\tilde\tau_R}}}{\frac{1}{\tilde\tau_A}+\frac{m^2}{\tilde\tau_R}}+\frac{1-e^{-\frac{\tilde\tau}{\tilde\tau_A}}}{\frac{1}{\tilde\tau_A}+\frac{n^2}{\tilde\tau_R}}+\frac{e^{-\frac{\tilde\tau}{\tilde\tau_A}}-e^{-\frac{m^2\tilde\tau}{\tilde\tau_R}}}{\frac{1}{\tilde\tau_A}-\frac{m^2}{\tilde\tau_R}}\right]
\end{align}
with $\tilde\tau_R \equiv (2M+1)^2/(\pi^2\tilde k)$ the dimensionless Rouse relaxation time.

\subsection{Analytical expression of $\langle \tilde \eta \circ \dot{\tilde x}_0 \rangle^{\rm ss}$}
\label{sec:work_rouse}

Consider the coordinate transformation from the original one $\{\tilde x_{-M},\tilde x_{-M+1},\cdots,\tilde x_{M}\}$ to its normal mode $\{\check x_{0},\check x_{1},\cdots, \check x_{2M}\}$ associated with Eq.~(\ref{eq:MSDeom}) with $U^{\rm int}({\bm x}) = \frac{k}{2}\sum_{i=-M}^{M-1}(\tilde x_{i+1}-\tilde x_{i})^2$. Their relationship can be written as~\cite{foldes2021assessing} 
\begin{align}
    \tilde x_n &=\check{x}_0 +2\sum_{m=1}^{2M}\check{x}_m \cos{\bigg[\frac{m\pi(n+M+\frac{1}{2})}{2M+1}\bigg]},\nonumber \\
    \check{x}_n &=\frac{1}{2M+1}\sum_{m=-M}^{M}\tilde {x}_m \cos{\bigg[\frac{n\pi(m+M+\frac{1}{2})}{2M+1}\bigg]}.
\label{eq:r_normalmode}
\end{align} 
This transformation is based on the following orthogonality relation:
\begin{align}
    &\sum_{l=-M}^M \cos{\bigg[\frac{m\pi(l+M+\frac{1}{2})}{2M+1}\bigg]} \cos{\bigg[\frac{n\pi(l+M+\frac{1}{2})}{2M+1}\bigg]} \nonumber \\
    &= \frac{(2M+1)\delta_{m,n}}{2-\delta_{m,0}} .
\end{align}
We can also write the same transformations for the variables $\tilde \xi_n$, $\tilde \eta_n$, and $\tilde F_n$. Especially, using the relations $\tilde \eta_n = \delta_{n,0} \eta$ and $\tilde F_n = \delta_{n,0} F$, the transformed coordinates of $\tilde \eta_n$ and $\tilde F_n$ can be simply expressed as 
\begin{align}
    &\check \eta_n = \frac{1}{2M+1}\tilde \eta \cos\left(\frac{n\pi}{2} \right), \nonumber \\
    &\check F_n = \frac{1}{2M+1}\tilde F \cos\left(\frac{n\pi}{2} \right).
\end{align}
Plugging these normal mode expansions of $\tilde x_n$, $\tilde \xi_n$, $\tilde \eta_n$, and $\tilde F_n$ into Eq.~(\ref{eq:MSDeom}) leads to the diagonalized normal mode equations as follows:
\begin{equation}
    \dot{\check{x}}_n = -k_n \check{x}_n + \check{F}_n +\check{\eta}_n  +\check{\xi}_n ,
    \label{eq:eom_rouse_normalmode}
\end{equation}
where $k_n \equiv 4\tilde k \sin^2{\big(\frac{n\pi}{4M+2}\big)}$. In addition, we can show that the autocorrelation functions for 
$\check{\xi}_n(t)$ and $\check{\eta}_n(t)$ are 
\begin{align}
    &\langle \check{\xi}_m( \tilde t)\circ \check{\xi}_n(\tilde t')\rangle=
    \frac{1}{2N+1}\delta_{m,n}(1+\delta_{m,0})\delta(\tilde t-\tilde t'), \nonumber \\
    &\langle \check{\eta}_m(\tilde t)\circ \check{\eta}_n(\tilde t')\rangle=\frac{\tilde v_p^2}{(2M+1)^2}e^{-|\tilde t-\tilde t'|/\tilde \tau_A}\cos{\Big(\frac{m\pi}{2}\Big)\cos{\Big(\frac{n\pi}{2}}\Big)}.
    \label{eq:corr_active_normalmode}
\end{align}
The exact solution of Eq.~(\ref{eq:eom_rouse_normalmode}) is 
\begin{equation}
    \check{x}_n(\tilde t)=\int_{-\infty}^{\tilde t} d\tilde t' e^{-{k_n (\tilde t-\tilde t')}}\left[\check{F}_n +\check{\eta}_n(\tilde t') + \check{\xi}_n(\tilde t')\right].
    \label{eq:eom_rouse_normalmode_solution}
\end{equation}
Using the normal mode expansion, we can express $\langle \tilde\eta\circ\dot{\tilde x}_0\rangle^{\rm ss}$ as the sum of $\langle\check{\eta}_m\circ\dot{\check{x}}_n\rangle^{\rm ss}$:
\begin{align} \label{eqA:ActiveEnergyCost}
    \langle \tilde\eta\circ\dot{\tilde x}_0\rangle^{\rm ss}
    =&
    \langle \check{\eta}_0\circ\dot{\check{x}}_0\rangle^{\rm ss} \\
    &+2\sum_{m=1}^{2M} \left(\langle \check{\eta}_0\circ\dot{\check{x}}_m\rangle^{\rm ss}+\langle \check{\eta}_m\circ\dot{\check{x}}_0\rangle^{\rm ss}\right)\cos{\Big(\frac{m\pi}{2}\Big)} \nonumber \\
    &+4\sum_{m=1}^{2M}\sum_{n=1}^{2M}\langle \check{\eta}_m\circ\dot{\check{x}}_n\rangle^{\rm ss}\cos{\Big(\frac{m\pi}{2}\Big)}\cos{\Big(\frac{n\pi}{2}\Big)}.
\end{align} 
From Eqs.~(\ref{eq:eom_rouse_normalmode}), (\ref{eq:corr_active_normalmode}), and (\ref{eq:eom_rouse_normalmode_solution}), $\langle \check{\eta}_m \circ\dot{\check{x}}_n \rangle^{\rm ss}$ can be evaluated as 
\begin{equation} \label{eqA:transformed_eta_dotx}
    \langle \check{\eta}_m \circ\dot{\check{x}}_n \rangle^{\rm ss}
    =\frac{ \tilde v_p^2}{(2M+1)^2}\frac{1}{\tilde k_m\tilde\tau_A+1}\cos{\Big(\frac{m\pi}{2}\Big)\cos{\Big(\frac{n\pi}{2}}\Big)}.
\end{equation} 
Inserting Eq.~(\ref{eqA:transformed_eta_dotx}) into Eq.~(\ref{eqA:ActiveEnergyCost}) with the relation $\sum_{n=1}^{2M}\cos^2\left(\frac{n\pi}{2}\right)=M$ finally leads to
\begin{equation}
    \langle \tilde\eta \circ\dot{ \tilde x}_0 \rangle^{\rm ss} =
    \frac{ \tilde v_p^2}{2M+1}\Bigg(1+2\sum_{m=1}^{2M}\frac{\cos^2{\Big(\frac{m\pi}{2}\Big)}}{ k_m\tilde\tau_A+1}\Bigg).
\end{equation}
Therefore, the energy consumption by the active noise is always positive and proportional to the square of the propulsion velocity of AOUP.

\bibliography{ref.bib}